\magnification\magstephalf
\overfullrule 0pt
\input epsf.tex

\font\rfont=cmr10 at 10 true pt
\def\ref#1{{\hbox{\rfont {\ [#1]}}}}


\font\fourteenbf=cmbx12 scaled\magstep1

\font\tenbfit=cmbxti10
\font\sevenbfit=cmbxti10 at 7pt
\font\fivebfit=cmbxti10 at 5pt
\newfam\bfitfam 
\textfont\bfitfam=\tenbfit  \scriptfont\bfitfam=\sevenbfit
\scriptscriptfont\bfitfam=\fivebfit

  \def\g {\gamma} \def \d {\delta}
\def\e{\epsilon}  \def\o{\omega}
  \def\la{\lambda}
  
\def\pd {\partial}
\def\pmb#1{\setbox0=\hbox{#1}
 \kern.05em\copy0\kern-\wd0 \kern-.025em\raise.0433em\box0 }

\def\bphi{\vec\phi}
\def\slash{/\kern-.5em}

\def \half {{\textstyle {1 \over 2}}}

 %


\def\boxit#1{\vbox{\hrule\hbox{\vrule\kern1pt\vbox
{\kern1pt#1\kern1pt}\kern1pt\vrule}\hrule}}

\def\h{\hfill\break}
\parskip=6pt
\parindent=0pt
\hsize=17truecm\hoffset=-5truemm
\vsize=23truecm
\def\footnoterule{\kern-3pt
\hrule width 17truecm \kern 2.6pt}


\catcode`\@=11 

\def\nolabels{\def\wrlabeL##1{}\def\eqlabeL##1{}\def\reflabeL##1{}}
\def\writelabels{\def\wrlabeL##1{\leavevmode\vadjust{\rlap{\smash%
{\line{{\escapechar=` \hfill\rlap{\sevenrm\hskip.03in\string##1}}}}}}}%
\def\eqlabeL##1{{\escapechar-1\rlap{\sevenrm\hskip.05in\string##1}}}%
\def\reflabeL##1{\noexpand\llap{\noexpand\sevenrm\string\string\string##1}}}
\nolabels
\global\newcount\refno \global\refno=1
\newwrite\rfile
\def\defref{{{\hbox{\rfont \ [\the\refno]}}}\nref}
\def\nref#1{\xdef#1{\the\refno}\writedef{#1\leftbracket#1}%
\ifnum\refno=1\immediate\openout\rfile=refs.tmp\fi
\global\advance\refno by1\chardef\wfile=\rfile\immediate
\write\rfile{\noexpand\item{#1\ }\reflabeL{#1\hskip.31in}\pctsign}\findarg}
\def\findarg#1#{\begingroup\obeylines\newlinechar=`\^^M\pass@rg}
{\obeylines\gdef\pass@rg#1{\writ@line\relax #1^^M\hbox{}^^M}%
\gdef\writ@line#1^^M{\expandafter\toks0\expandafter{\striprel@x #1}%
\edef\next{\the\toks0}\ifx\next\em@rk\let\next=\endgroup\else\ifx\next\empty%
\else\immediate\write\wfile{\the\toks0}\fi\let\next=\writ@line\fi\next\relax}}
\def\striprel@x#1{} \def\em@rk{\hbox{}} 
\def\lref{\begingroup\obeylines\lr@f}
\def\lr@f#1#2{\gdef#1{\defref#1{#2}}\endgroup\unskip}
\newwrite\lfile
{\escapechar-1\xdef\pctsign{\string\%}\xdef\leftbracket{\string\{}
\xdef\rightbracket{\string\}}}

\def\writestop{\def\writestoppt{\immediate\write\lfile{\string\p
ageno%
\the\pageno\string\startrefs\leftbracket\the\refno\rightbracket%
\string\def\string\secsym\leftbracket\secsym\rightbracket%
\string\secno\the\secno\string\meqno\the\meqno}\immediate\closeout\lfile}}
\def\writestoppt{}\def\writedef#1{}
\catcode`\@=12 
%
\def\0{\over } \def\1{\vec } \def\2{{1\over2}} \def\4{{1\over4}}
\def\5{\bar } 
\def\6{\partial }
\def\7#1{{#1}\llap{/}}
\def\8#1{{\textstyle{#1}}} \def\9#1{{\bf{#1}}}
\def\.{\cdot }
\def\^#1{\widehat{#1}}
\def\({\left(} \def\){\right)} \def\<{\langle } \def\>{\rangle }
\def\[{\left[} \def\]{\right]}

\rightline{\tt hep-ph/9708426}
\rightline{DAMTP 97/74}
\rightline{TUW 97-15}
\vskip 4mm
\centerline{\fourteenbf FOAM DIAGRAM SUMMATION}
\medskip
\centerline{\fourteenbf AT FINITE TEMPERATURE}
\vskip 6mm
\centerline{\bf I T Drummond, R R Horgan, P V Landshoff}
\vskip 1mm
\centerline{DAMTP, University of Cambridge$^\star$}
\vskip 3mm
\centerline{\it and}
\vskip 3mm
\centerline{\bf A Rebhan}
\vskip 1mm
\centerline{Institut f\"ur Theoretische Physik, Technische Universit\"at Wien,
Vienna$^\star$}
\footnote{}{$^\star$ itd@damtp.cam.ac.uk \  rrh@damtp.cam.ac.uk \
pvl@damtp.cam.ac.uk \ rebhana@tph16.tuwien.ac.at}
\vskip 0.5 cm
{\bf Abstract}

We show that large-$N$ $\phi ^4$ theory is not trivial if one accepts the
presence of a tachyon with a truly huge mass, and that it allows exact
calculation. We use it to illustrate how to  calculate the exact resummed
pressure at finite temperature and verify that it is infrared and ultraviolet
finite even in the zero-mass case.
In 3 dimensions a residual effect of the resummed infrared divergences
is that at low temperature or strong coupling
the leading term in the interaction pressure becomes 
independent of the coupling and is 4/5 of the free-field pressure. 
In 4 dimensions the pressure is well defined
provided that the temperature is below the tachyon mass.
We examine how  rapidly this expansion converges and use our analysis to
suggest how one might reorganise perturbation theory to improve the
calculation of the pressure for the QCD plasma.
\vskip 10mm

{\bf 1 Introduction}

In traditional calculations of the pressure in massless quantum
field theories at finite temperature, there is
a breakdown of perturbation theory because of infrared problems\defref\kap{
J I Kapusta, {\sl Finite Temperature Field Theory}, Cambridge University
Press (1989);\h
T Altherr, Physics Letters B238 (1990) 360}\defref\lebellac{
M Le Bellac, {\sl Thermal Field Theory}, Cambridge University Press (1996)
}. 
This is true, in particular, of gauge theories at finite temperature.
Formally the same types of problem crop up also in scalar 
$\lambda\phi^4$ theory. The traditional procedure\ref{\kap} for
summing the ring diagrams
(figure 1a) produces a series expansion for the pressure $P$ in powers
(and logarithms) of $\lambda ^{1/2}$. 
This takes care of infinitely many otherwise problematic higher-loop
diagrams, 
but it does not cover diagrams that include a single one-particle-irreducible
self-energy insertion and so leaves 
arbitrarily high-loop orders to be considered.
In nonabelian gauge theories, where some propagators remain massless
perturbatively, the two-particle-irreducible higher-loop diagrams are
still potentially infrared divergent.
In an earlier paper\defref\us{
I T Drummond, R R Horgan, P V Landshoff and A Rebhan, Phys Lett B398 (1997) 326
}
we have proposed a different and less direct method which absorbs all of
the diagrams into a single resummed one-loop quantity
which enjoys manifest infrared regularity in four dimensions. 
This introduces  a variable mass $m$ for the field as a parameter, and
gives the pressure in terms of an integral over the thermal propagator
corresponding to mass $m$. 

In this paper, we apply our summation method to 
sum all the extended ring diagrams
of figure 1b, which we call foam diagrams. Other authors
\defref\DJ{L Dolan and 
R Jackiw, Phys Rev D9 (1974) 3320}
\defref\ACP{G Amelino-Camelia and S-Y Pi, Phys Rev D47 (1993) 2356;\h
G Amelino-Camelia, hep-ph/9702403}
have called these super-daisy or  Hartree-Fock diagrams.
We began this analysis as a warm-up exercise for the gauge-theory problem,
but have found that it has interesting features in its own right.

Moreover, it allows us to investigate the convergence properties of
thermal perturbation theory in a solvable model, 
from which we derive suggestions on how
to optimize the perturbative treatment of the pressure also in more
complicated theories such as QCD.
\topinsert
\line{{\hfill\epsfxsize=40truemm\epsfbox{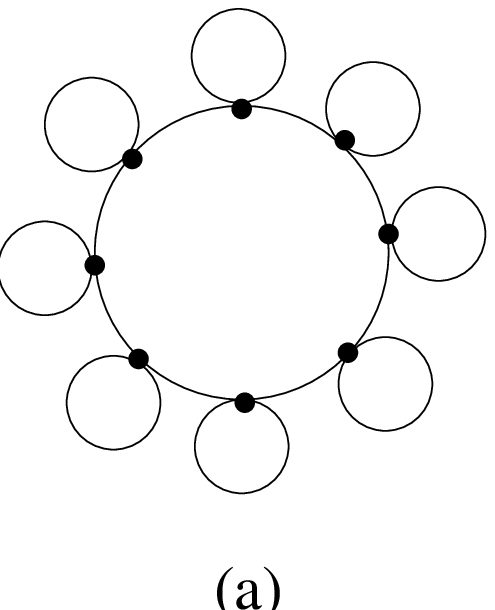}}\hfill
{\epsfxsize=100truemm\epsfbox{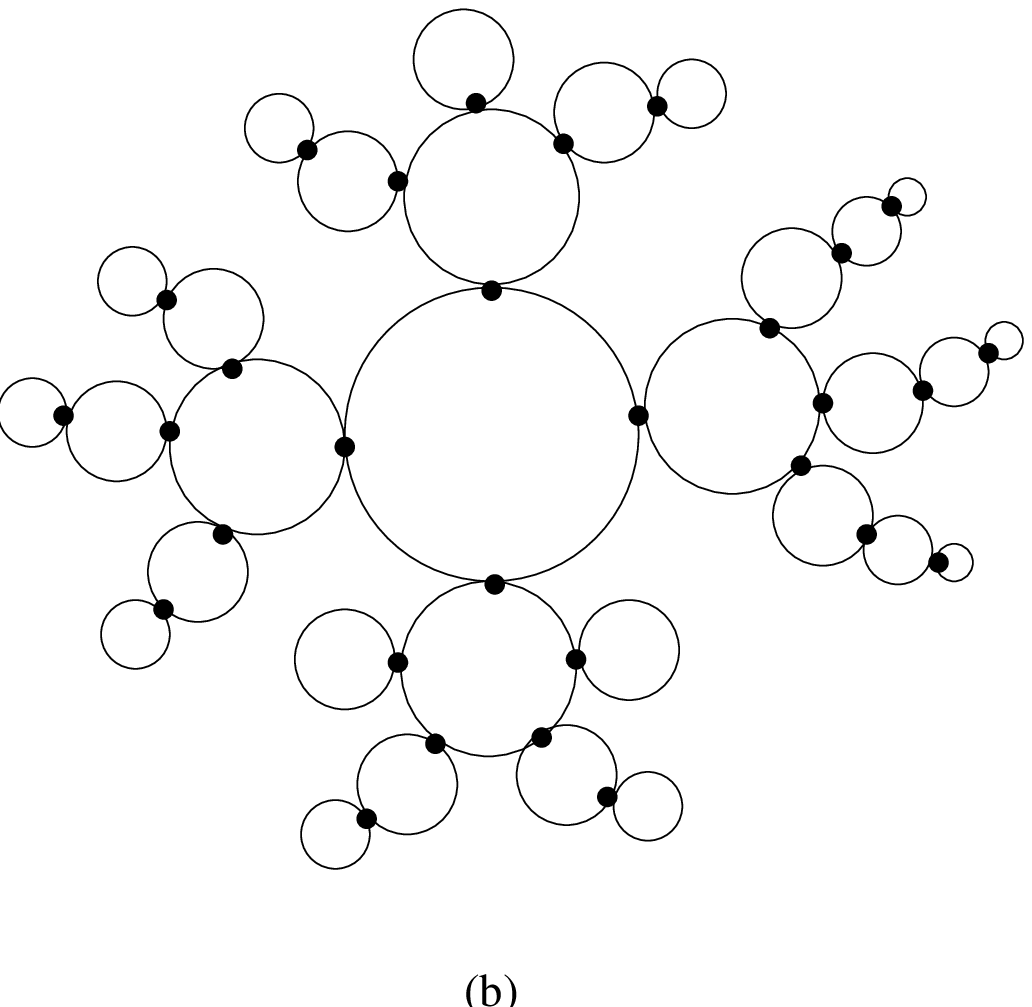}}\hfill}\hfill\break
\centerline{Figure 1: (a) ring diagrams; (b) foam diagrams}
\endinsert

If we 
consider an $O(N)$ scalar field theory with unrenormalised Lagrangian density
$$
{\cal L}(x)=\half\left( (\pd{{\bphi }}(x))^2-m_0^2 ({{\bphi }}(x))^2\right )
-{\lambda _0\over 4!}{3\0N+2} 
\left (({{\bphi }}(x))^2\right )^2
\eqno(1.1)
$$
then in the limit $N\to\infty$ the pressure per scalar particle 
coincides with the one obtained from the infinite sum of foam diagrams in
the $N=1$ theory. Hence our calculation in  this paper of the foam diagrams
may be regarded either as the leading-$N$ term in the pressure in the
large-$N$ theory, or as an approximation (we do not know how  good) to
the pressure in the $N=1$ theory. 
We obtain an expression for this infinite sum that
is derived from  a path integral and so has validity beyond perturbation
theory.  The renormalisation of the mass, and in 4 dimensions also of the
coupling, is an essential
part of the analysis; this is achieved through Dyson equations which can be
solved exactly, and so again goes beyond perturbation theory.

We describe the renormalisation in section 2. As is well known\defref\trivial{
K G Wilson and J Kogut, Phys Rep 12 (1974) 75;\h
D J E Callaway, Phys Rep 167 (1988) 241;\h
J Zinn-Justin, {\sl Quantum Field Theory and Critical Phenomena}, 
Oxford Science Publications (1996)
}, and as we find, in 4 dimensions there are problems with $\phi ^4$ theory. 
For reasonably small coupling,
these problems turn out to arise only at huge mass scales and so are not really
important for physics, but in order to handle them 
we write our equations for $n$ dimensions. In section 3 we review our
formula\ref{\us} for the pressure; as before, we choose to use the real-time
formalism of thermal field theory.  The formula involves the thermal addition
$\delta m^2$ to the renormalised squared mass $m^2$.
A key simplification is that, for the foam diagrams, $\delta m^2$ is both
independent of momentum  and real, and we show how this leads to rather simpler
versions of our general formula for the pressure. We need to solve an integral
equation for $\delta m^2$, which we do in section 4 for the case of 3
dimensions. 
We show that the infrared divergences that occur in the zero-mass case
are rendered harmless by our resummation and that in the strong-coupling
or low-temperature limit
the leading term in the interaction pressure becomes
independent of the coupling and is 4/5 of the free-field pressure.

In section 5 we investigate what happens in 4 dimensions
if we ignore the triviality difficulties of the theory. We find that
the integral equation for  $\delta m^2$ has no solution above some critical
temperature $T_{\hbox{\sevenrm max}}$,
so that it seems that then the
pressure does not exist. However, for
sufficiently weak coupling, $T_{\hbox{\sevenrm max}}$ is
exponentially huge, so that 
the 4-dimensional theory may be accepted as a highly accurate
effective theory.
In the particularly interesting massless case we make a detailed comparison
of our nonperturbative results with those of resummed perturbation theory.
We find that the latter converges quickly only for rather small coupling
and that the rate of convergence depends critically on the renormalization
scale. We end with some speculation on the case of QCD, where explicit
calculation of the pressure up to and including order $g^5$ has revealed
rather bad convergence properties of resummed perturbation theory there. 
We suggest that it may well be fruitful to reorganise the expansion so
that $\d m/T$, rather than the coupling $g$, is the expansion parameter.
\bigskip
{\bf 2 Renormalisation}
\medskip
{\it Zero temperature}

We choose a zero-temperature renormalisation scheme that makes the
formula for the pressure
$P(T)$ as simple as possible. In lowest-order perturbation theory
the renormalised mass is defined to be
$$
m^2=m_0^2+\la _0M(m_0^2)
\eqno(2.1a)
$$
where $M$ corresponds to the single-loop Feynman graph:
$$
M(m^2)=\half\int {d^nq\over(2\pi)^n}{i\over q^2-m^2+i\e}
\eqno(2.2a)
$$
We extend this to all orders of perturbation theory by replacing (2.1a)
with the Dyson equation
$$
m^2=m_0^2+\la _0M(m^2)
\eqno(2.1b)
$$
The integration over $q^0$ in the integral in (2.2a) may be done by closing
the contour in one or
other half plane and taking the residue at the pole. The result is that
instead we may write 
$$
M(m^2)=\half\int {d^nq\over(2\pi)^n}2\pi\delta ^+(q^2-m^2)
\eqno(2.2b)
$$
The integration is simple:
$$
M(m^2)={\Gamma(1-\half n)\over 2(4\pi )^{n/2}}m^{n-2}
\eqno(2.2c)
$$

When we go to 4 dimensions we need also to renormalise the coupling. Initially
we keep $n\not=4$ as a regulator.
We define the renormalised coupling $\lambda$ to be the value of the $ii\to jj$
scattering amplitude at $s=0$, where $i$ and $j$ denote ``colour'' labels.
The equation for $\lambda$ is shown diagrammatically in figure 2.
If we work to leading order in $N$ in the large-$N$ theory, or choose to
include only foam diagrams in the $N=1$ case,
we must omit the last two terms in figure 2,
so that we have
$$
\la=\la _0+\la _0\la   L(m^2)$$$$
L(m^2)=\half\int {d^nq\over(2\pi)^n}{i\over (q^2-m^2+i\e )^2}= M'(m^2)
\eqno(2.3a)
$$
where the prime denotes differentiation with respect to $m^2$.
Both $M$ and $M'$ are ultraviolet divergent when $n\to 4$.
The coupling has dimension $m^{4-n}$. From (2.3a) and (2.2c),
$$
\la={\la _0\over 1+C_n\la _0m^{n-4}}$$$$
C_n={\Gamma (2-\half n)\over 2(4\pi )^{n/2}}
\eqno(2.4a)
$$
so that if we want both the bare and the renormalised
coupling to be non-negative
$$
0\leq \la \leq {m^{4-n}\over C_n}
\eqno(2.5)
$$
Hence, when $n\to 4$, $\la$ vanishes for all values of $m$,
because $C_n$ diverges. This is the well-known triviality of $\phi ^4$
theory in 4 dimensions\ref{\trivial}.

\topinsert
\centerline {{\epsfxsize=90truemm\epsfbox{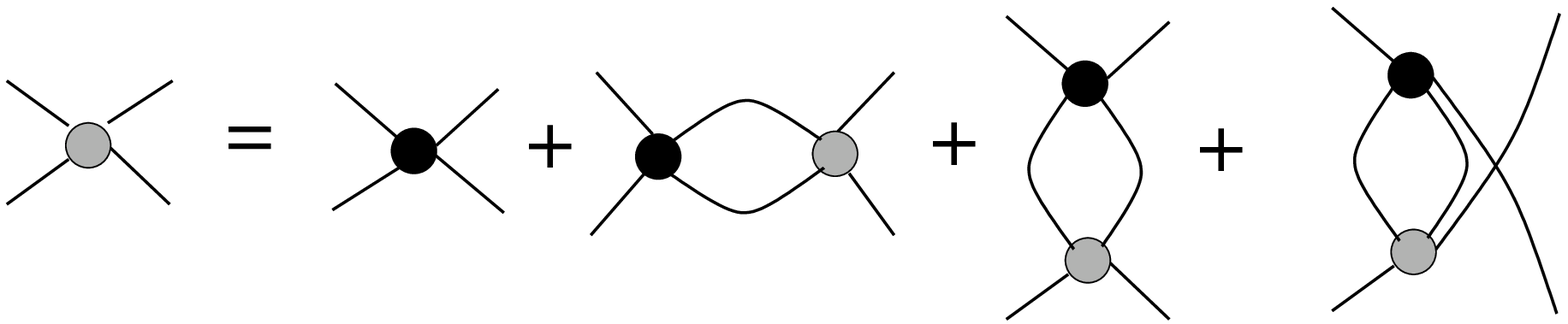}}}\hfill\break
\centerline{Figure 2: coupling renormalisation}
\endinsert

One might perhaps believe that it does not matter whether the bare coupling
is positive, because only the renormalised theory is relevant. However,
if for $n=4$ we choose $\la$ to be greater than 0, the renormalised
theory has a tachyon.  To see this, we write the equation corresponding
to figure 2 (with the last two terms omitted because they are negligible
in the large-$N$ limit)
for the $ii\to jj$ scattering amplitude for a general value of $s$:
$$
T(s)=\la _0 +\la _0 M'(m^2, s) T(s)$$$$\eqalignno{
M'(m^2,s)&=-\half\int {d^nq\over (2\pi )^n}
{i\over \left ((q+\half p)^2-m^2+i\epsilon\right )
        \left ((q-\half p)^2-m^2+i\epsilon\right )}\cr
   &=-C_n\int _0^1dx \left (m^2-sx(1-x)\right )^{n/2 -2}
&(2.6)\cr}
$$
{}From (2.4a) and (2.6),
$$
T(s)={\la\over 1+C_n\la
\int _0^1dx \{\left ( m^2-sx(1-x)\right )^{n/2 -2}-m^{n-4}\}}
\eqno(2.7a)
$$
which becomes in 4 dimensions when $|s|$ is large
$$
T(s)\approx {\la\over 1-{\la\over 32\pi ^2}\log (-s/m^2)}
\eqno(2.7b)
$$
It is evident that, in 4 dimensions, $T(s)$ has a pole at 
$$
s=s_{\hbox{\sevenrm tachyon}}\approx-m^2 \exp \left ({32\pi^2\over\la}\right )
\eqno(2.8a)
$$
Our definition (2.3a) of the renormalised coupling $\la$ in terms of
the value of the scattering amplitude $T(s)$ at $s=0$ makes $\la$ vary with
the renormalised mass $m$, and the appearance of a 
tachyon pole in $T(s)$
is related to the occurrence of a Landau pole in $\la (m^2)$.
Although this tachyon pole is in principle unacceptable, in practice 
it occurs at a very large negative value of $s$. For example, if $\la=1$,
it is at $-10^{137}m^2$. One could avoid the presence of this tachyon by
introducing an UV cut-off $\Lambda$, since then
$$
\la={\la_0 \over 1+{\la_0\over 32\pi^2}\log(\Lambda^2/m^2)}
\eqno(2.4b)
$$
which makes it possible to have both $\la$ and $\la_0$ positive,
with $\la < 32\pi^2/\log(\Lambda^2/m^2)$ and therefore 
$|s_{\hbox{\sevenrm tachyon}}|> \Lambda^2$.

Or alternatively we simply say that, for reasonably small coupling, the
tachyon is so far away
that it can be ignored. Either way, the $n=4$ theory seems to be perfectly
acceptable as an effective theory, for energies much below either the
tachyon mass or the cut-off.

In what follows we shall be particularly interested in the case of
massless theories, because there one encounters the infrared problems
of thermal perturbation theory. Because of the tachyon 
(2.8a), this case may seem to be excluded, for then the tachyon
cannot be kept far away, unless we send $\la\to0$ as $m\to0$.
Contrary to appearances, this does not signal a trivial theory, however.
In the massless case, there are infrared divergences in the defining
equation for the renormalised coupling (2.3a). Switching to an alternative
scheme defined by
$$
\bar\la=\la_0+\la_0\bar\la L(\bar\mu^2)		
\eqno(2.3b)
$$
with some arbitrary scale $\bar\mu$,
shows that then, when $\bar\la\not=0$,
$$
T(s)={\bar\la\over 1-{\bar\la\over 32\pi ^2}\left[\log (-s/\bar\mu^2)-2\right]}
\eqno(2.7c)
$$
vanishes at $s=0$, but not otherwise.
So whenever we are going to inspect the massless limit, we shall
switch over to using $\bar\la$ rather than $\la$.
In terms of $\bar\la$ and $\bar\mu$, the tachyon is at
$$
s_{\hbox{\sevenrm tachyon}}=-\bar\mu^2 \exp \left ({32\pi^2\over\bar\la}
+2 \right )
\eqno(2.8b)
$$
so for given choice of $\bar\mu$
we can always make it as far away as we wish by taking $\bar\la$
small enough.

In minimally subtracted massless theories, a renormalization scale
is usually introduced through $\lambda\to\mu^{4-n}\lambda$. Because
$$
\mu^{4-n} L(\bar\mu^2)\sim -{1\over32\pi^2}\( {2\over4-n}-\gamma-\log{\bar\mu^2
\over 4 \pi \mu^2} \)
$$
the scale $\bar\mu^2$ introduced as in (2.3b) is seen to coincide with
the one of the modified minimal subtraction scheme ($\overline{\hbox{MS}}$), 
$\bar\mu^2=
4\pi e^{-\gamma}\mu^2$.
Hence our notation.

A more physical renormalization scheme would be to define the
renormalized coupling through the scattering amplitude at some
scale $\widetilde\mu^2$ so that
$\widetilde\lambda=T(-\widetilde\mu^2)$.
(2.7c) shows that $\widetilde\mu^2\equiv e^2\bar\mu^2$.
However, in conformity with most of the literature on thermal field
theory, we shall continue to use the $\overline{\hbox{MS}}$ scheme.

\goodbreak

\medskip
{\it Nonzero temperature}

At nonzero temperature, in addition to
the renormalised zero-temperature mass there is a 
thermal contribution, because (2.1) becomes replaced with
$$
m^2+\delta m^2=m_0^2+\la _0 M_T(m^2+\delta m^2)$$$$
M_T(m^2)=M(m^2)+N_T(m^2)$$$$
N_T(m^2)=\int {d^nq\over(2\pi)^n}2\pi\delta ^+(q^2-m^2){1\over e^{q^0/T}-1}
\eqno(2.9)
$$
Eliminating the unrenormalised mass $m_0^2$ with (2.1b) gives 
$$
\delta m^2(m^2,T)=\la _0 [M_T(m^2+\delta m^2)-M(m^2)]
\eqno(2.10)
$$
For less than 4 dimensions both $\la _0$ and the expression in square
brackets are finite, but when $n\to 4$ this is no longer true and
we must introduce also the renormalised coupling.
If we define the function
$$\eqalignno{
\hat M(m^2,\delta m^2)&=M(m^2+\delta m^2)-M(m^2)-\delta m^2M'(m^2)\cr
&=C_n\left\{\big (m^2+\delta m^2\big )^{n/2-1}-(m^2)^{n/2-2} \big (
m^2+(n/2-1)\delta m^2\big  )\right\}&(2.11a)
\cr}
$$
we find that 
$$
\delta m^2(m^2,T)=\la [\hat M(m^2,\delta m^2)
+N_T(m^2+\delta m^2)]
\eqno(2.12)
$$
This is an integral equation for $\delta m^2(m^2,T)$ in which
everything is finite when $n\to 4$, with
$$
\hat M(m^2,\delta m^2)\to {1\over 32\pi ^2}\left\{(m^2+\delta m^2)
\log \left (1+{\delta m^2\over m^2}\right )-\delta m^2\right\}
\eqno(2.11b)
$$
The expansion of this in powers of $\delta m^2/m^2$ begins with a 
term proportional to $(\delta m^2)^2$.
In perturbation theory $\delta m^2=O(\la)$, so that 
$\la\hat M=O(\la^3)$ and the first term on the right-hand side of
(2.12) represents a contribution
that first appears at three loop order. 
Since frequently calculations up to two loops
have been taken as a clue to exact results, this particular contribution
has been repeatedly missed in the literature\ref\DJ
\defref\AOS{J Arafune, K Ogure and J Sato, hep-th/9705158}.
It owes its existence to the interplay 
between zero-temperature contributions plus 
their renormalisation with the thermal effects, and it
will turn out to be of great importance to the existence and
behaviour of the solutions of the 
equation for $\delta m^2$.

In the massless limit $m\to0$, we are again confronted with infrared
divergences, which can be avoided by switching to the alternative
renormalisation scheme (2.3b) This amounts to substituting
$$
\la^{-1}=\bar\la^{-1}+{1\032\pi^2}\log{\bar\mu^2\0m^2}
\eqno(2.13)
$$
which, in the limit $m\to0$, leads to
$$
\delta m^2(0,T)=\bar\la [\bar M(\bar\mu^2,\delta m^2)
+N_T(\delta m^2)]
\eqno(2.14)
$$
with
$$
\bar M(\bar\mu^2,\delta m^2)=
{1\032\pi^2}\delta m^2\;(\log{\delta m^2\0\bar\mu^2}-1)
\eqno(2.15)
$$
It is easy to see 
that $\delta m^2(0,T)$ is, as it should  be, 
independent of $\bar\mu$ once the $\bar\mu$-dependence
of $\bar\la$ is taken into account. 

\bigskip
{\bf 3 The pressure}

Our formula\ref{\us}  for the pressure is derived from the grand partition
function 
$$
Z(T)=\int d\bphi\, \exp(iS[\bphi,T])$$$$
S[\bphi,T])=\int _Cd^nx\,{\cal L}(x)
\eqno(3.1)
$$
In the variant of the real-time thermal field theory we use\footnote{$^*$}%
{For a review of the essentials of real-time thermal field theory,
see reference\defref\pvl{
P  V Landshoff, {\sl Introduction to thermal field theory},  lectures at
Andr\'e Swieca Summer School, hep-ph/9705376
}},
the integration is over all {\bf x} and over the time contour $C$ which,
in the complex $t$-plane, runs along the real axis from $-\infty$ to
$+\infty$, back to $-\infty$, and then down to $-\infty -i/T$.
(This is known as the Keldysh contour\defref\keldysh{
L V Keldysh, Sov Phys JETP 20 (1965) 1018
}.) Differentiate with respect to $m_0^2$ keeping $\la _0$ fixed:
$$\eqalignno{
{\pd\over\pd m_0^2}\log Z &=
-\half iZ^{-1}\int  d\bphi\,\left ( \int _C d^nx\, \bphi ^2(x)\right  )
\exp (iS[\bphi ,T] )\cr
&=-\half i\big < \int _C d^nx\, \bphi ^2(x)\big >
&(3.2)\cr}
$$
Here, $<\dots >$ denotes a thermal average.
Space-time translation invariance tells us that the thermal average of
$\bphi ^2(x)$ is independent of $x$, and so the $x$ integration is trivial:
$$\eqalignno{
{\pd\over\pd m_0^2}\log Z&=-{V\over 2T}\big <  \bphi^2 (0)\big >\cr
&=-{NV\over 2T}\int {d^nq\over(2\pi)^n}\, D^{12}(q,T)
&(3.3)\cr}
$$
Here,
$$
D^{12}(q,T)=\left <\int d^nx\,e^{iq.x}\,
\phi _i(0)\phi _i(x)\right >
\eqno(3.4)
$$
where $\phi _i$ denotes any component of the $O(N)$-symmetric field $\bphi$.
$D^{12}(q,T) $ is an element of the familiar $2\times 2$ matrix propagator
${\bf D}(q,T)$ of Keldysh-contour
real-time thermal field theory\ref{\lebellac}\ref{\pvl}. 
Inserting its known form\defref\lvw{
N P Landsman and Ch G van Weert,
Physics Reports 145 (1987) 142}\defref\pvlar{
P V Landshoff and A Rebhan, Nucl Phys B410 (1993) 23
}
into (3.3) and using the relation $P(T)=(T/V)\log Z$, we obtain
$$
{\pd P(T)\over \pd m_0^2}=
\int {d^nq\over(2\pi)^n}
{\epsilon (q^0 )\over e^{q^0 /T}-1}\;
\hbox{Im }{1\over q^2-m_0^2-\Pi (q,T,m)}
\eqno(3.5a)
$$

When, as in the application to the foam diagrams, the self energy $\Pi$
is real, we must apply the usual $i\epsilon$ prescription to $m_0^2$
and so
$$\eqalignno{
- {\pd P(T)\over\pd m_0^2}&=M_T(m_0^2+\Pi (q,T,m))\cr 
&=M_T(m^2+\delta m^2)
&(3.5b)\cr}
$$
where $M_T$ is defined in (2.9).
Subtracting off the zero-temperature pressure and using (2.10), we have
$$
{\pd\over\pd m_0^2}\big (P(T)-P(0)\big )
=-{\delta m^2(m^2,T)
\over\lambda _0}
\eqno(3.6)
$$
Using (2.1a) and (2.3a) to express everything in terms of
renormalised quantities and imposing the obvious boundary condition that the
pressure vanishes for infinite mass, we find the remarkable formula
$$
P(T)-P(0)=\int_{m^2}^\infty dm'^2{\delta m^2(m'^2,T)
\over\lambda(m'^2)} 
\eqno(3.7a)
$$
Here, 
$$
\lambda(m'^2)={\lambda\over 1+\la [L(m^2)-L(m'^2)]}
\eqno(3.7b)
$$
so that $\lambda(m'^2)$ is a running coupling that is equal to $\la$ when
$m'^2=m^2$; it is calculated from (2.3a) by varying the mass and keeping
fixed $\la _0$.
\def\M{{\cal M}}
\def\N{{\cal N}}

In fact, the mass integration in (3.7a) can be carried out.
Rewriting the right-hand side of (3.5b) as
$$\eqalign{
 M_T(m_0^2+\Pi (q,T,m)) &= M_T(m_0^2+\Pi)\;\Big( 1+{\6\Pi\0\6m_0^2}\Big)-
{1\0\la_0}\Pi{\6\Pi\0\6m_0^2} \cr
&=M_T(m^2+\delta m^2){\6\0\6m_0^2}(m^2+\delta m^2)
-{1\02\la_0}{\6\0\6m_0^2}\Pi^2\cr}
\eqno(3.8a)
$$
and introducing
$$
\M(m^2)=\int _{m^2}^{\infty}dm'^2\,M(m'^2)=
\half\int {d^nq\over(2\pi )^{n-1}}\theta (q^0)\theta (q^2-m^2)$$$$ 
\N_T(m^2)=\int _{m^2}^{\infty}dm'^2\,N_T(m'^2)=
\int {d^nq\over(2\pi )^{n-1}}\theta (q^0)\theta (q^2-m^2){1\over e^{q^0/T}-1}
\eqno(3.8b) 
$$
we have
$$
M_T(m_0^2+\Pi (q,T,m))=
-{\6\0\6m_0^2}\[ \M_T(m^2+\delta m^2)+{1\02\la_0}\Pi^2(T,m) \]
$$
so that with the boundary condition that the pressure vanishes for
infinite mass
$$\eqalign{
P(T)-P(0)&=\M_T(m^2+\delta m^2)-\M(m^2)+{1\02\la_0}\(\Pi^2(T,m)-\Pi^2(0,m)\)\cr
&=\N_T(m^2+\delta m^2)+\M(m^2+\delta m^2)-\M(m^2) + 
\half\delta m^2\big (
M_T(m^2+\delta m^2)+M(m^2)\big ) \cr}
\eqno(3.9)
$$
A formula similar to (3.9) 
has been derived previously by Amelino-Camelia and Pi\ref{\ACP} using the
CJT formalism\defref\CJT{J M
Cornwall, R Jackiw and E Tomboulis, Phys Rev D10 (1974) 2428}, though 
their formula does not satisfy the physically-important constraint
that the pressure vanishes when the mass is infinite.\footnote{$^*$}{In
reference \ref\CJT, the effective potential is calculated, from which
the pressure of our model follows by restricting to vanishing field
expectation value and positive bare mass squared. We intend considering the
symmetry-breaking sector of the theory in a future paper.}

In order to highlight
the interplay between thermal and quantum contributions,
let us also give the following alternative version of the result (3.9)
$$
P(T)-P(0)=\N_T(m^2+\delta m^2)+ \half {(\d m^2)^2\0 \la}-
\sum_{n=3}^\infty {1\0n!}M^{(n-1)}( m^2)(\d m^2)^n
\eqno(3.10)$$
This expression makes manifest the UV finiteness\footnote{$^{**}$}{Its
IR finiteness is however better seen from the original version
(3.9).} of our result, and it
exhibits three different kinds of contribution:
$\N_T$, the classical expression for the
pressure of a bosonic gas of particles with mass squared $m^2+\d m^2$;
$\;\;(\d m^2)^2/ \la$, which is $O( \lambda )$ in perturbation theory,
essentially a thermal interaction contribution; and the rest, which
starts at three loop order, coming from the thermal mass shift in
zero-temperature integrals.
\goodbreak

\bigskip
{\bf 4 Three dimensions}

When $n=3$, (2.2c) and  (2.9)  become 
$$
M(m^2)=-{m\over 8\pi}
$$$$
N_T(m^2)=-{T\over 4\pi}\log (1-e^{-m/T})
\eqno(4.1)
$$
Hence the integral equation (2.10) for $\delta m^2$ reduces to
$$
{8\pi 
\over\la _0}\delta m^2=m-\sqrt{m^2+\delta m^2}
-2T\log \left (1-\exp (-\sqrt{m^2+\delta m^2}/T)\right )
\eqno(4.2)
$$
This must be solved numerically and the result plugged into (3.9).
(We recall that, in 3 dimensions, $\la _0$ has the dimension of mass.)

When $m=0$,
$$
{8\pi T \over\la _0}\xi ^2+\xi =
-2\log \left ( 1-\exp (-\xi)\right )
\eqno(4.3)
$$
where $\delta m^2=\xi ^2T^2$. 
The solution for $\xi$
becomes small at high temperature, 
when $8\pi T\gg\la _0$, but it goes to 0 quite slowly:
it is about 0.4 for $8\pi T/\la _0=10$ and 0.07 for $8\pi T/\la _0=1000$.

In terms of $\xi$, the formula for the pressure can be written as
$$
P(T)-P(0)={T^3\02\pi}\left\{ {\rm Li}_3(e^{-\xi})+\xi\, {\rm Li}_2(e^{-\xi})
-{\xi^2\04}\log(1-e^{-\xi})+{\xi^3\024} \right\}
\eqno(4.4)$$
where ${\rm Li}_3$ and ${\rm Li}_2$ are the tri- and di-logarithmic
functions, respectively\defref\Lewin{L Lewin, {\it Polylogarithms and
Associated Functions}, North Holland Publishing (1981)}. In the
high-temperature limit, where $\xi\to0$, this approaches the free-field
value $\zeta(3)T^3/(2\pi)$.

Notice that a perturbation expansion of the pressure in the $m=0$ case
would be fraught with infrared problems. These would be encountered
already at order $\la _0^2$, in the two-loop graph. The formula (3.9)
has eliminated them through resummation. The right-hand side of (3.9)
is finite when $m\to 0$, yielding (4.4),
but it does not have a power-series expansion
in powers of $\delta m^2$, because the derivative of $M_T$ diverges at
the origin.
This infrared sensitivity leads to the interesting result that  at
low temperature or strong coupling the 
interaction pressure becomes a constant multiple of the free-field pressure,
independent of the coupling.
For $8\pi T\ll\la _0$ the first term
in (4.3) becomes negligible compared with the other two,
and the solution for $\xi$ 
approaches
$$
\xi\equiv\delta m/T \to \log{3+\sqrt5\02} \qquad \hbox{for $T/\la_0 \to0$}
\eqno(4.5)$$
that is $\delta m\approx 0.96\,T$.

Remarkably, for this value of $\xi$ the polylogarithms in (4.4) can be
evaluated (see equations (1.20) and (6.13) of reference\ref{\Lewin}), yielding
$$
P(T)-P(0) \to {2 \zeta(3) \0 5\pi} T^3 = {4\05} [P(T)-P(0)]_{\rm free}
\qquad \hbox{for $T/\la_0 \to0$}
\eqno(4.6)$$

We have not found any physical explanation of this surprisingly simple
result nor have we been able to derive it without recourse to the
peculiar properties of polylogarithms.

\goodbreak

\bigskip
{\bf 5 Four dimensions}

\medskip
{\it  The massive case}

When $n\to 4$ we must use the integral equation (2.12) for $\delta m^2$,
which is written in terms of the renormalised coupling $\la$.
With $x=\delta m^2/m^2$ and $z=T/m$, it reads
$$
 \la=[F(x,z)]^{-1}$$$$
F(x,z)={1\over 32\pi ^2x}\Big\{ (1+x)\log (1+x)-x+8(1+x)
\int _1^{\infty} d\o {\sqrt{\o ^2-1}\over e^{\o\sqrt{1+x}/z}-1}\Big\}
\eqno(5.1a)
$$
The function $[F(x,z)]^{-1}$ is plotted against $x$ in figure 3,
for various values of $T/m$. Because of the factor $1/x$ in $F$,
all the curves go to 0 at small $x$. Because of the logarithm term in $F$,
which arises from the
(frequently neglected) contribution $\hat M$ in (2.11b), they go to
0  again at 
large $x$. So for each, there is a maximum choice of $\la$ beyond which
the equation has no real solution; for example, for $T/m=3$ the critical value
of $\la$ is about 200. 
For values of $\la$ below the critical value 
there are two solutions, as was found previously by
Bardeen and Moshe\defref\BM{
W A Bardeen and M Moshe, Phys Rev D28 (1983) 1372}.
However,  only the smaller one is relevant, since by 
definition $\delta m^2\to 0$ as $T\to 0$; 
the larger solution corresponds
rather to having a second, nonperturbative solution for the renormalized
mass $m$ as a function of the bare mass $m_0$.
For reasonable coupling 
the larger solution is in fact exponentially huge so that it would not
matter anyway.
For overcritical $\lambda$, the two solutions
become complex and have to be dismissed because we have started with
the assumption of a real self-energy $\Pi$ in (3.5b).
\midinsert
\centerline {{\epsfxsize=120truemm\epsfbox{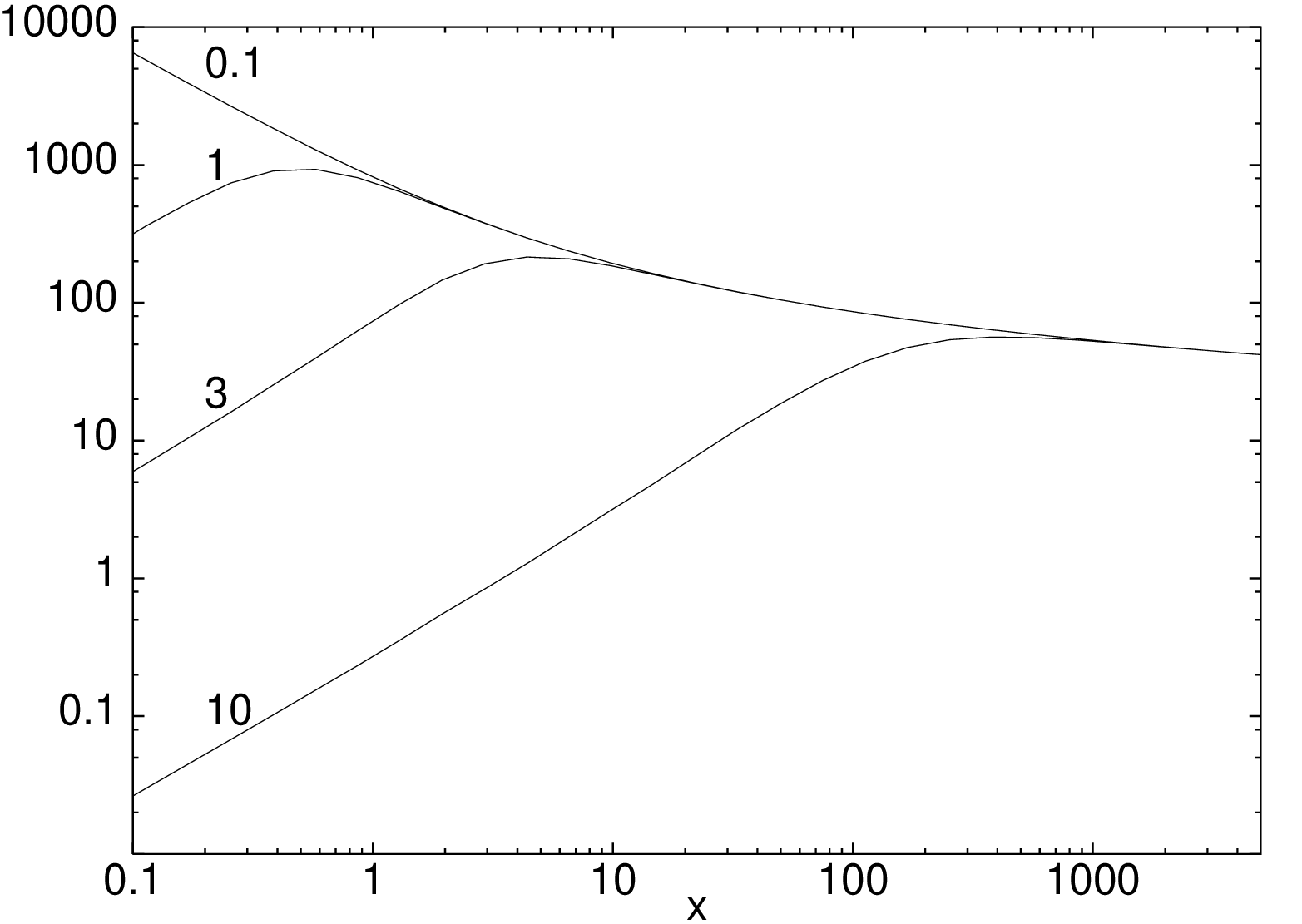}}}\hfill\break
\centerline{Figure 3: the function $F^{-1}$ in (5.1a) plotted against
$x$ for various values of $T/m$}
\endinsert

Alternatively, for a given choice of $\la$, there is a critical value
of $T/m$ beyond which there is no solution. For $\la<10$ this
critical value is large and we may find it from
(5.1a) approximately. The stationary value of the right-hand side of (5.1a)
occurs when
$$
{z\over 4\sqrt{x}}=\int _1^{\infty}dk{\sqrt{k^2-1}\over(e^{k\sqrt{x}/z}-1)
(1-e^{-k\sqrt{x}/z})}
\eqno(5.2)
$$
The integral is approximately 0.8 when $\sqrt{x}/z=1$ and it decreases rapidly
as $\sqrt{x}/z=1$ increases. Hence the stationary value is at $x$ just
greater than $z^2$, and so the critical value is given by 
$$
T\approx m\exp\left (16\pi ^2\over\la\right )
$$
That is, the critical temperature is of the order of the tachyon mass,
which for $\la=1$ is about $10^{68}m$. 

\medskip
{\it The massless case}

\topinsert
\centerline {{\epsfxsize=100truemm
\epsfbox{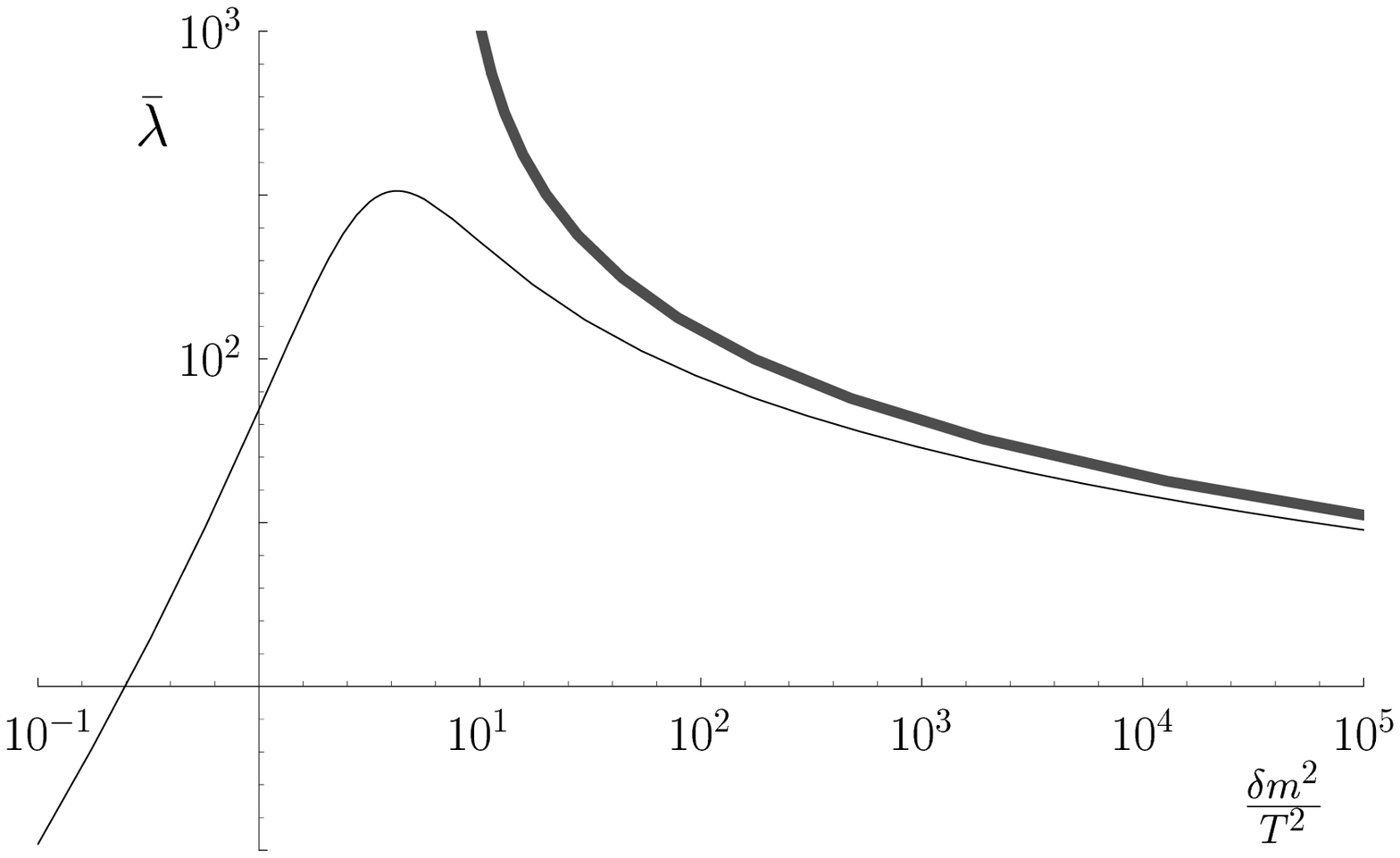}}}
\moveright1cm\vbox{\hsize=14cm
Figure 4: the function 
$\bar\la=\bar F^{-1}$ for $m=0$ plotted against
$\delta m^2/T^2$; the heavy line marks where $\delta m$ would
become equal to the (modulus of the) tachyon mass.}
\endinsert

As we have seen in section 2, the massless case requires that $\la\to0$
as $m\to0$. This does not mean that we are driven to a trivial theory,
but only that the definition of the coupling constant through the
scattering at zero energy becomes inappropriate. Instead
we have to introduce a renormalization scale $\bar\mu$ and a finite
coupling constant $\bar\la$ as given by (2.3b) or (2.13).
The integral equation to be solved for $\delta m^2$ is now (2.14).
In place of (5.1a) we have
$\bar\la=[\bar F(\delta m^2,\bar\mu)]^{-1}$
and the situation is similar to the one of the massive case with
$\bar\mu$ replacing $m$. If for example we choose $\bar\mu$ such
that $\bar\la(\bar\mu)=1$, the critical temperature above which there
are no solutions is of the order of $\bar\mu\exp(16\pi^2)$. 

In the remainder of this paper, we shall compare the exact results
that we have found against a perturbative evaluation.
At a given nonzero temperature it will turn out to be useful
to define the coupling at $\bar\mu=T$ through
$$
\bar\la (T)={\bar\la (\bar\mu )\over 1+\bar\la (\bar\mu )[L(\bar\mu)-L(T)]}
\eqno(5.3)
$$
With this choice of the renormalization scheme the solutions to
$\bar\la=[\bar F(\delta m^2,\bar\mu=T)]^{-1}$ are plotted in figure 4.
The critical value of $\bar\lambda(T)$ beyond which no solutions exist
turns out to be approximately 325.5.
Below this, there are two solutions for the
thermal mass, but the higher solution is found in the region
close to the tachyon scale which is marked by the heavy line in figure 4.
For $\bar\lambda(T) 
\ll 10^2$, the latter is exponentially far away
and this loosely defines the range of coupling where we can accept the theory
as an effective one. 

By the way, had we chosen $\bar\mu/T>1.62\ldots$ we would have
found a seemingly different picture: for those values of $\bar\mu$ 
the critical value $\bar\la(\bar\mu)$
has moved past infinity so that there always exist two solutions for
positive coupling; but this is only because this change of the
renormalization scheme maps sufficiently large values of 
$\bar\lambda(\bar\mu=T)$
onto negative $\bar\lambda(\bar\mu)$.
However, all this occurs in the region close to the
tachyon mass scale, which we can ignore by avoiding
too-large couplings.

We now compare the exact results one gets by
a numerical evaluation of the above formulae against a perturbative
one. This is particularly interesting in the case of a massless theory, for
there ordinary perturbation theory runs into infrared
singularities that need to be cured by resummation of the thermal
mass.

In perturbation theory $\d m^2/T^2 \sim \lambda$, but a naive expansion
of the functions $N_T$ and $\N_T$ as a power series in their argument
is bound to fail. However we may write the function $N_T(\d m^2)$ that
appears in the integral equation (2.14) for $\d m^2$ as
$$N_T(\d m^2)={T^2\04\pi^2}\sum_{k=1}^\infty {\d m\0kT} K_1(k\,\d m/T)     
$$
where $K_1$ is the modified Bessel function of the second kind.
Using a Mellin transform\defref\Braden{H W Braden, Phys Rev D25 (1982) 1028}
one finds
$$\eqalign{
\sum_{k=1}^\infty {\d m\0kT}\, K_1\left ({k\,\d m\0T}\right )=&
{\pi^2\06}-{\pi \,\d m\02T}
-\4\left ({\d m\0T}\right )^2\[\log{\d m\04\pi T}+\g-\2\]\cr&
-\4\left ({\d m\0T}\right )^2\sum_{n=1}^\infty (-1)^n 
{(2n)!\0(n+1)!\,n!}\,\zeta(2n+1)
\({\d m\04\pi T}\)^{2n}\cr}                                            
\eqno(5.4)$$
With (5.4) we can solve the equation 
for the thermal mass
perturbatively to any
desired accuracy in $\bar\la$ and insert the
result into the pressure (3.9), which in the massless case is given by
$$
P(T)-P(0)=\N_T(\delta m^2)+\half\delta m^2 N_T(\delta m^2)+{1\0128\pi^2}
\delta m^4
\eqno(5.5)$$
For this, we need also
$$\eqalign{
{4\pi^2\0T^4}\N_T(\d m^2)=&
{2\pi^4\045}-{\pi^2\06}\left ({\d m\0T}\right )^2+
{\pi\03}\left ({\d m\0T}\right )^3
+{1\08}\left ({\d m\0T}\right )^4\[\log{\d m\04\pi T}+\g-{3\04}\]\cr&
+{1\04}\left ({\d m\0T}\right )^4\sum_{n=1}^\infty 
(-1)^n {(2n)!\0(n+2)!\,n!}\zeta(2n+1)
\({\d m\04\pi T}\)^{2n}\cr}                                            \eqno(5.6)$$
The infrared divergences that appear in conventional perturbation theory
without resummation of thermal masses have their origin in the fact that
the series (5.4) and (5.6) are nonanalytic in $\delta m^2$ at $\delta m=0$.

The first few terms of the series expansion in $\bar\la$ are
$$
{\delta m^2\0T^2}={\bar\la \over 24}-{\bar\la ^{3/2}\over 16\pi \sqrt{6}}
+(3-\g-\log{\bar\mu\04\pi T}){\bar\la^2\0384\pi^2}
-(1-2\g-2\log{\bar\mu\04\pi T}){\bar\la^{5/2}\01024\pi^3\sqrt{2/3}}
+O(\bar\la^3)
\eqno(5.7)$$
$$\eqalign{
{P(T)-P(0)\0T^4}=&{\pi^2\090}-{\bar\la\01152}+{\bar\la^{3/2}\0 576\pi\sqrt6}
-(6-\gamma-\log{\bar\mu\04\pi T}){\bar\la^2\018432\pi^2}
+(3-2\gamma-2\log{\bar\mu\04\pi T}){\bar\la^{5/2}\012288\pi^3\sqrt6}\cr&
-\( (6-\gamma-\log{\bar\mu\04\pi T})^2-30+{\zeta(3)\036} \)
{\bar\la^3\0294912\pi^4}
+O(\bar\la^{7/2})\cr}
\eqno(5.8)$$
It is simple to verify that (5.8) is independent of $\bar\mu$: differentiate
it with respect to $\log\bar\mu$ and use $d\bar\la/d\log\bar\mu 
=\bar\la ^2/16\pi^2$ from (2.13).

\pageinsert
\centerline {{\epsfxsize=97truemm
\epsfbox{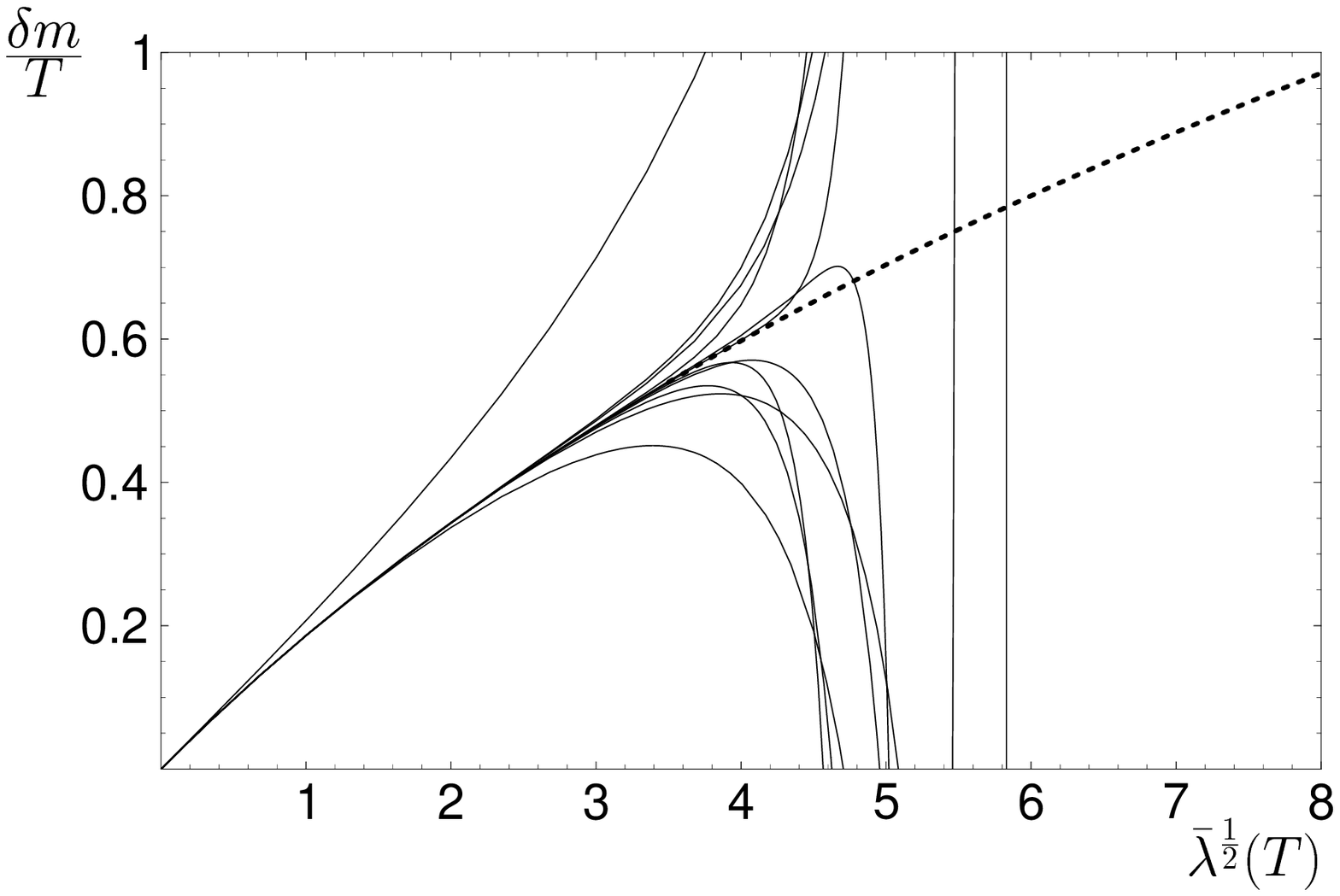}}}
\centerline{{(a)}}
\medskip
\centerline{{\epsfxsize=97truemm
\epsfbox{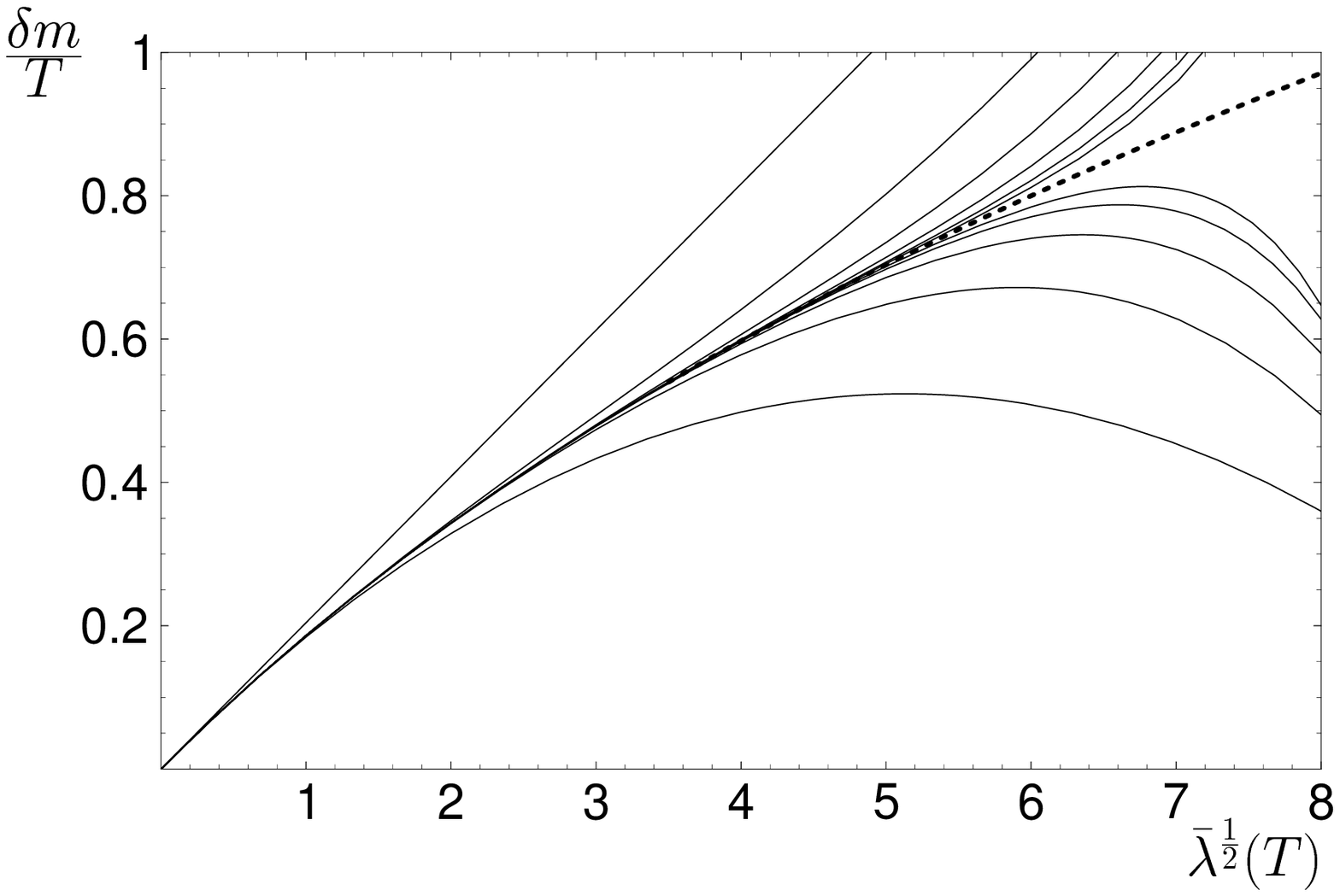}}}
\centerline{{(b)}}
\medskip
\centerline{{\epsfxsize=97truemm
\epsfbox{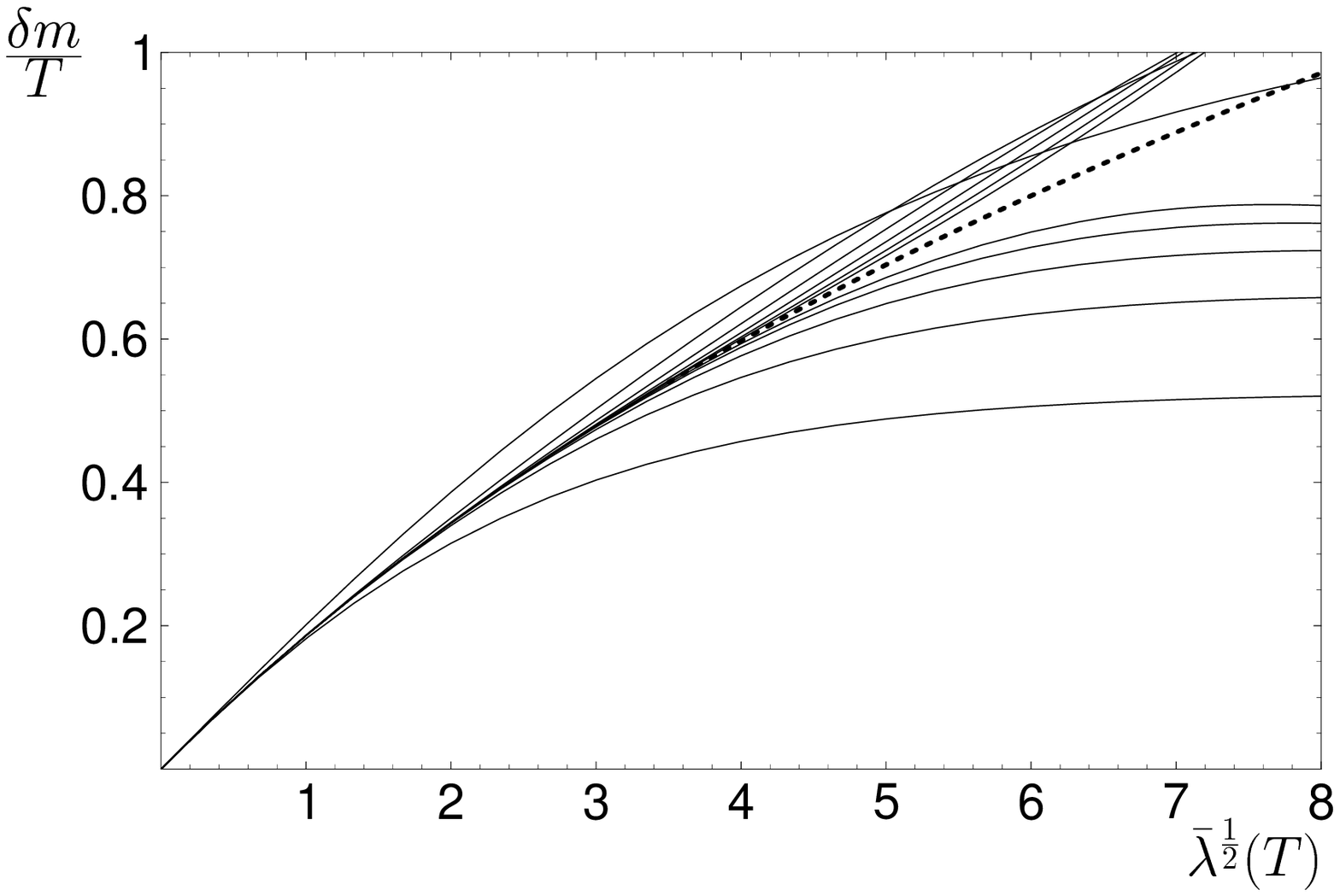}}}
\centerline{{(c)}}
\moveright1cm\vbox{\hsize=14cm
Figure 5: A comparison of the perturbative results for
$\delta m/T$ as a function of ${\bar\la^{1/2}(T)}$ 
up to 10th order for different choices of the
renormalization scale: a) $\bar\mu=100T$, b) $\bar\mu=T$, 
c) $\bar\mu={1\0100}T$. In a) the ``vertical'' lines are part of two of
the curves on the left.
}
\endinsert

\pageinsert
\centerline {{\epsfxsize=97truemm
\epsfbox{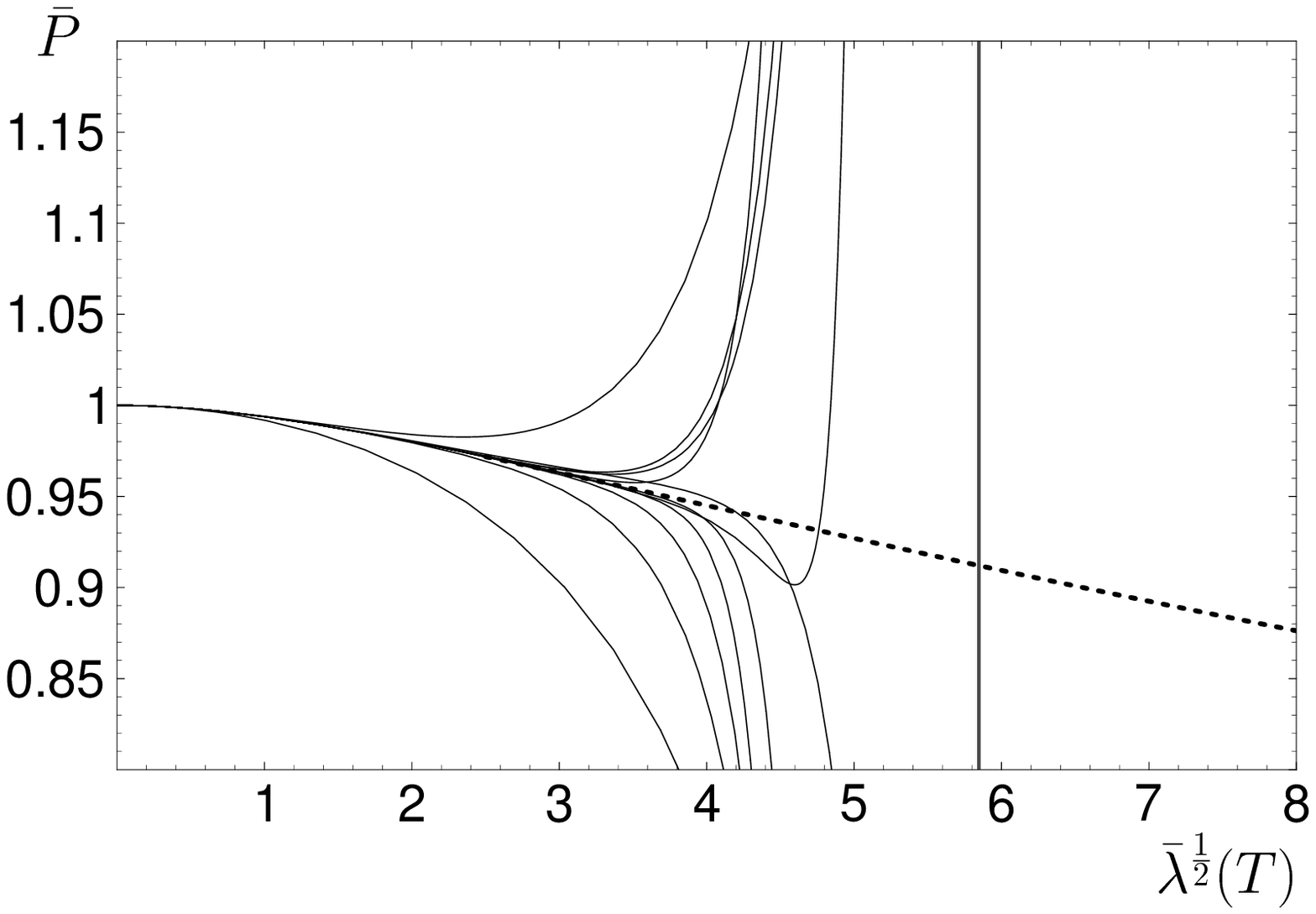}}}
\centerline{(a)}
\medskip
\centerline{{\epsfxsize=97truemm
\epsfbox{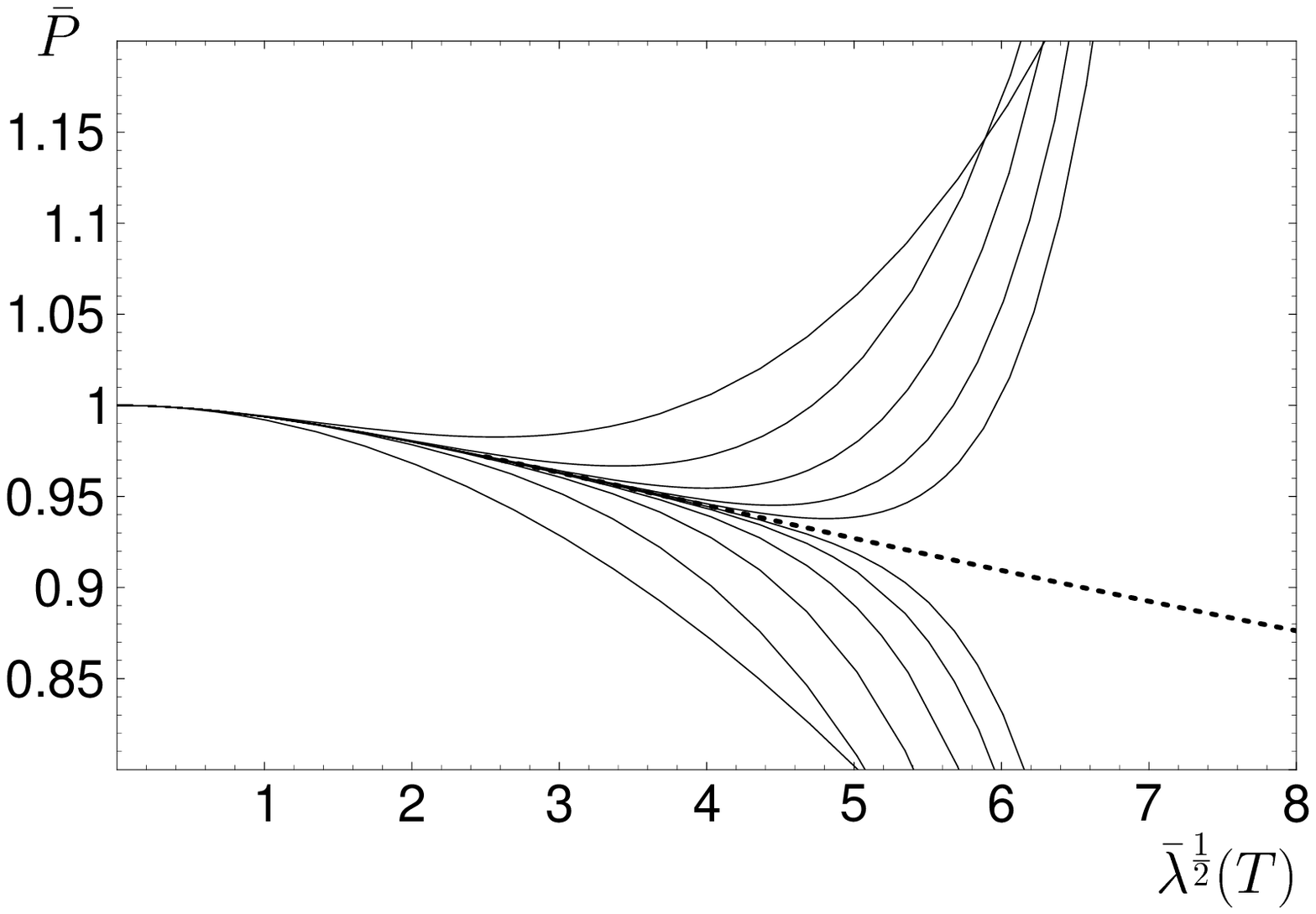}}}
\centerline{(b)}
\medskip
\centerline{{\epsfxsize=97truemm
\epsfbox{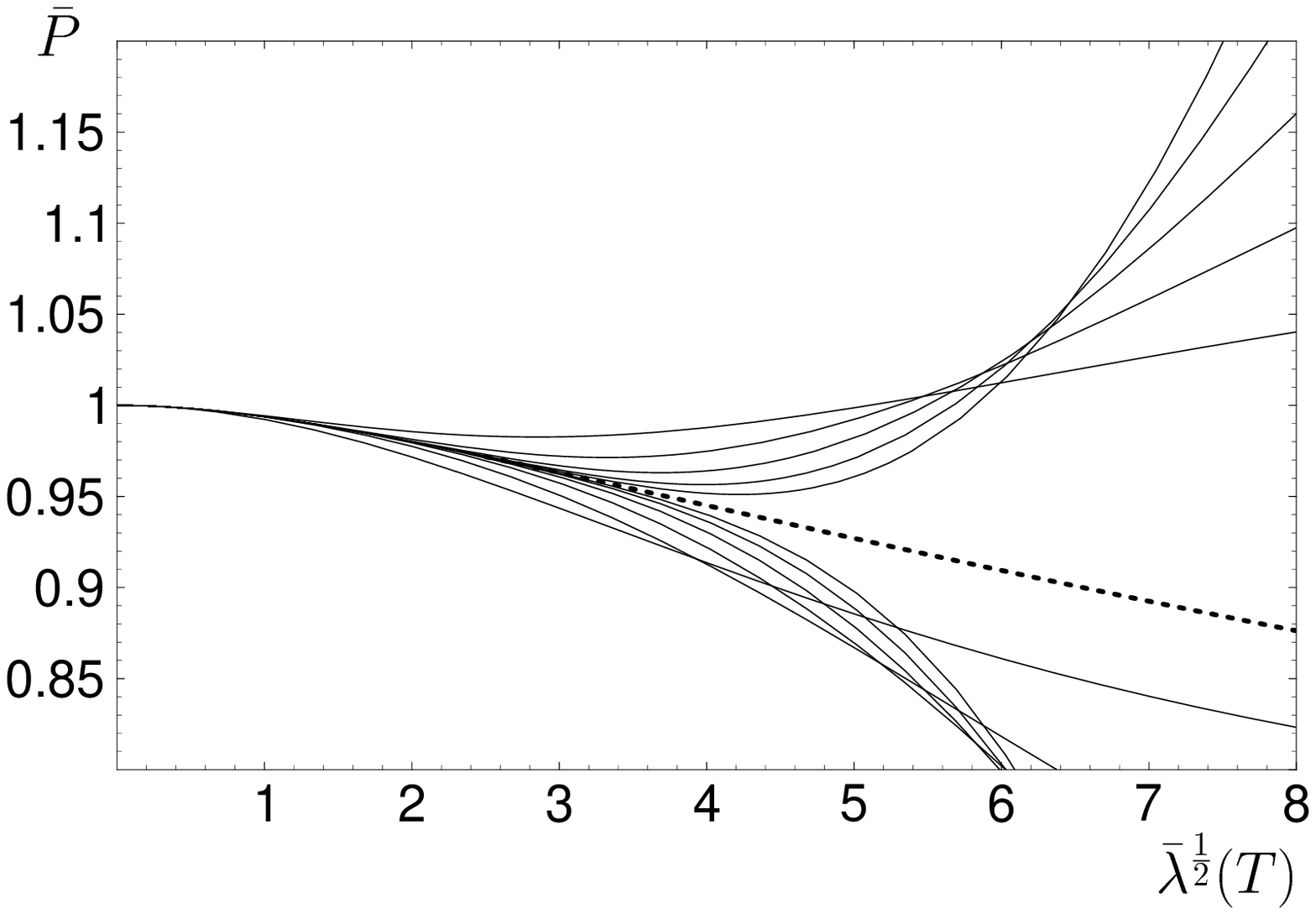}}}
\centerline{(c)}
\moveright1cm\vbox{\hsize=14cm
Figure 6: A comparison of the perturbative results for $\bar P\equiv
[P(T)-P(0)]/{\pi^2T^4\090}$
as a function of ${\bar\la^{1/2}(T)}$
up to 12th order for different choices of the
renormalization scale: a) $\bar\mu=100T$, b) $\bar\mu=T$, 
c) $\bar\mu={1\0100}T$.
}
\endinsert

\topinsert
\centerline{{\epsfxsize=95truemm
\epsfbox{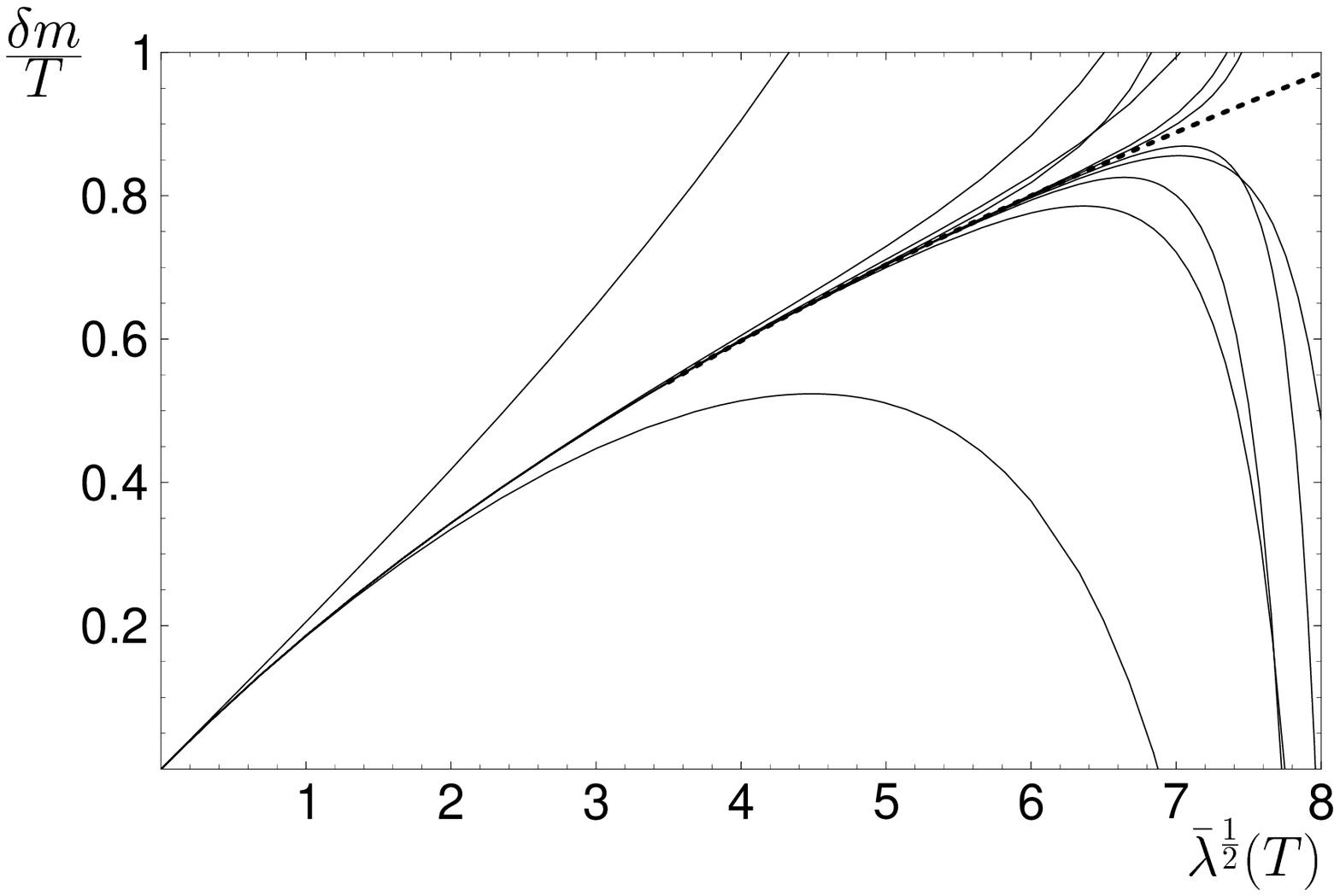}}}
\centerline{(a)}
\medskip
\centerline{\epsfxsize=95truemm
\epsfbox{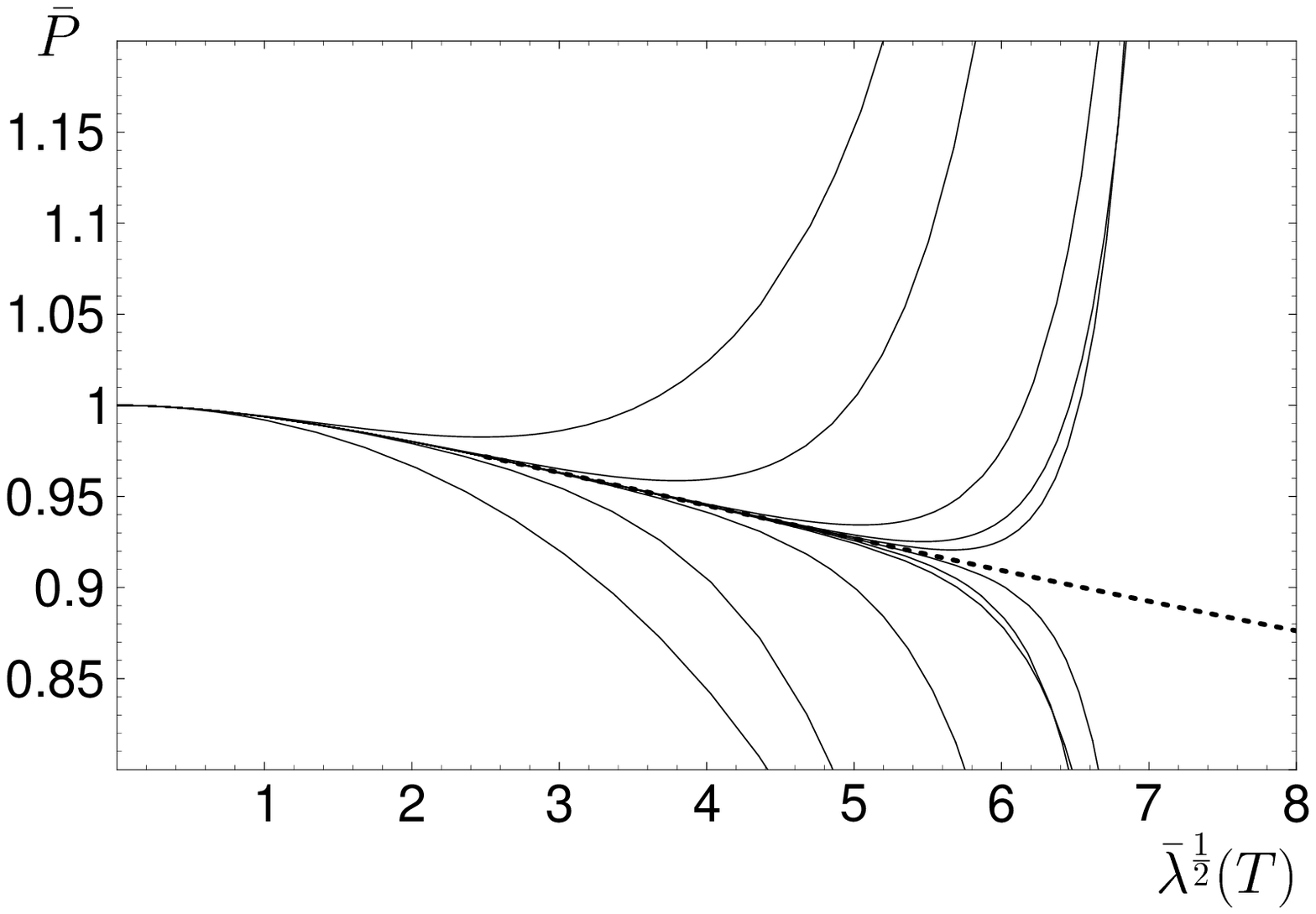}}
\centerline{(b)}
\medskip

\moveright1cm\vbox{\hsize=14cm
Figure 7: The perturbative results for $\delta m/T$ and 
$\bar P\equiv
[P(T)-P(0)]/{\pi^2T^4\090}$ for $\bar\mu=2\pi T$. 
Compared to figures 5b
and 6b, the rate of convergence becomes much more rapid after the first
few approximations, for which $\bar\mu=T$ is slightly favoured.}
\endinsert

In the case of the $N=1$ theory, the pressure has been calculated up to and 
including order 
$\bar\la^{5/2}$ using hard-thermal-loop resummed perturbation theory
\defref\PS{R R Parwani and H. Singh, Phys Rev D51 (1995) 4518}. Up to
and including order $\bar\la^{3/2}$, there is no difference between
the subset of foam diagrams and the complete set of diagrams, and the
results indeed agree. Beyond this order, 
there are differences, in particular there are no terms involving
$\log\bar\la$ in the foam-diagram
subset, which do occur in the full set starting at order $\bar\la^{5/2}$.
They come from the logarithmic terms in the expansion (5.4) and (5.6),
which in the case of foam diagrams happen to combine with the log in
(2.15) such that the thermal mass drops 
out from the arguments of the logarithms.

Despite this simplification, a comparison of the above perturbative
result with the exact one 
might give some hints about the convergence properties of thermal
perturbation series in general. 
It turns out that these depend strongly on the ratio of $\bar\mu/T$.

In figures 5-7 we juxtapose the exact and the perturbative
results for the thermal mass $\delta m/T$ and the ratio  of the pressure 
\hbox{$[P(T)-P(0)]$}
to its ideal-gas value $\pi ^2T^4/90$,
including in (5.7) and (5.8) up to 10 terms beyond the
leading one. We choose various 
values of the
renormalization scale $\bar\mu$,  but 
for ease of comparison
in each case we plot against $\bar\la$
evaluated for $\bar\mu=T$ through the relation (5.3).
(The actual expansion parameter $\bar\la(\bar\mu)$ is larger (smaller)
when $\bar\mu$ is larger (smaller) than $T$.)
The resulting 11 approximants are put on top of each other in order to
give a visual impression of the rate (or failure) of convergence of
the perturbative expansions in $\bar\lambda$; the exact results
are indicated by dashed lines.

When $\bar\mu$ is very different from $T$, the convergence of the series
deteriorates markedly. In figures 5 and 6 the results for the thermal mass
and the pressure are seen 
to become oscillatory for larger
coupling when $\bar\mu=100T$ 
(the vertical lines in figures 5a and 6a are part of these oscillating
results),
whereas with $\bar\mu={1\0100}T$ the perturbative results fail to improve
with higher orders at roughly the same place, although in a less
violent manner.

With the choice $\bar\mu=2\pi T$, which has been advocated in reference
\defref\BN{E Braaten and A Nieto, Phys Rev D 51
(1995) 6990}
on the grounds that this is the mass of the first nonzero
Matsubara mode, the behaviour of the series expansion 
can be significantly improved, although for the first few approximants
the results are slightly worse than at $\bar\mu=T$, see figure 7.

Attempting to improve perturbation theory by putting to zero one of the
$\bar\lambda^2$-terms in (5.7) or (5.8) gives much larger $\bar\mu$'s
and rather bad convergence properties. 
The optimal choice of $\bar\mu$ seems to be around 
$\bar\mu=4\pi \exp(-\gamma) T$, which absorbs all $\gamma$'s and
$\log(4\pi)$'s. This is in fact rather close to the choice $\bar\mu=2\pi T$
of reference \ref{\BN}.
Notice that the origin of the $\gamma$'s in (5.7) and (5.8) is entirely
from the high-temperature expansions (5.4) and (5.6); 
those appearing in dimensional regularization
have already been absorbed in $\bar\mu$.
However, without the restriction to foam diagrams, the $\gamma$'s would not
have the same coefficients as the logs, so it is not clear
whether 
this result for the optimal renormalization point could be a general one.
But it does confirm the expectation\ref{\BN} that the optimal renormalization
scheme is to be found around $2\pi T$ rather than $T$.

Another noteworthy observation is that the rate of convergence is
markedly slower for the perturbation series of the pressure than it is for 
the thermal mass.
Since this loss of accuracy comes from having inserted a perturbative
result for $\delta m/T$ into the series (5.6) and truncated at a
given order in the coupling, a certain improvement would be simply to
refrain from doing a high-temperature expansion
of the integrals that appear in the expression for the pressure.
The quality of the perturbation series for the pressure is then
the same as that of the thermal mass.
It would be interesting to see whether this could be implemented
in QCD to ameliorate
the frustratingly bad apparent convergence of resummed perturbation
theory for the QCD pressure, 
which has been calculated up to order $g^5$ recently\defref\QCDP{P Arnold
and C Zhai, Phys Rev D50 (1994) 7603; D51 (1995) 1906;\h
C Zhai and B Kastening, Phys Rev D52 (1995) 7232;\h
E Braaten and A Nieto, Phys Rev Lett 76 (1996) 1417; Phys Rev D53 (1996) 3421}.

\topinsert
\centerline {
{\epsfxsize=95truemm
\epsfbox{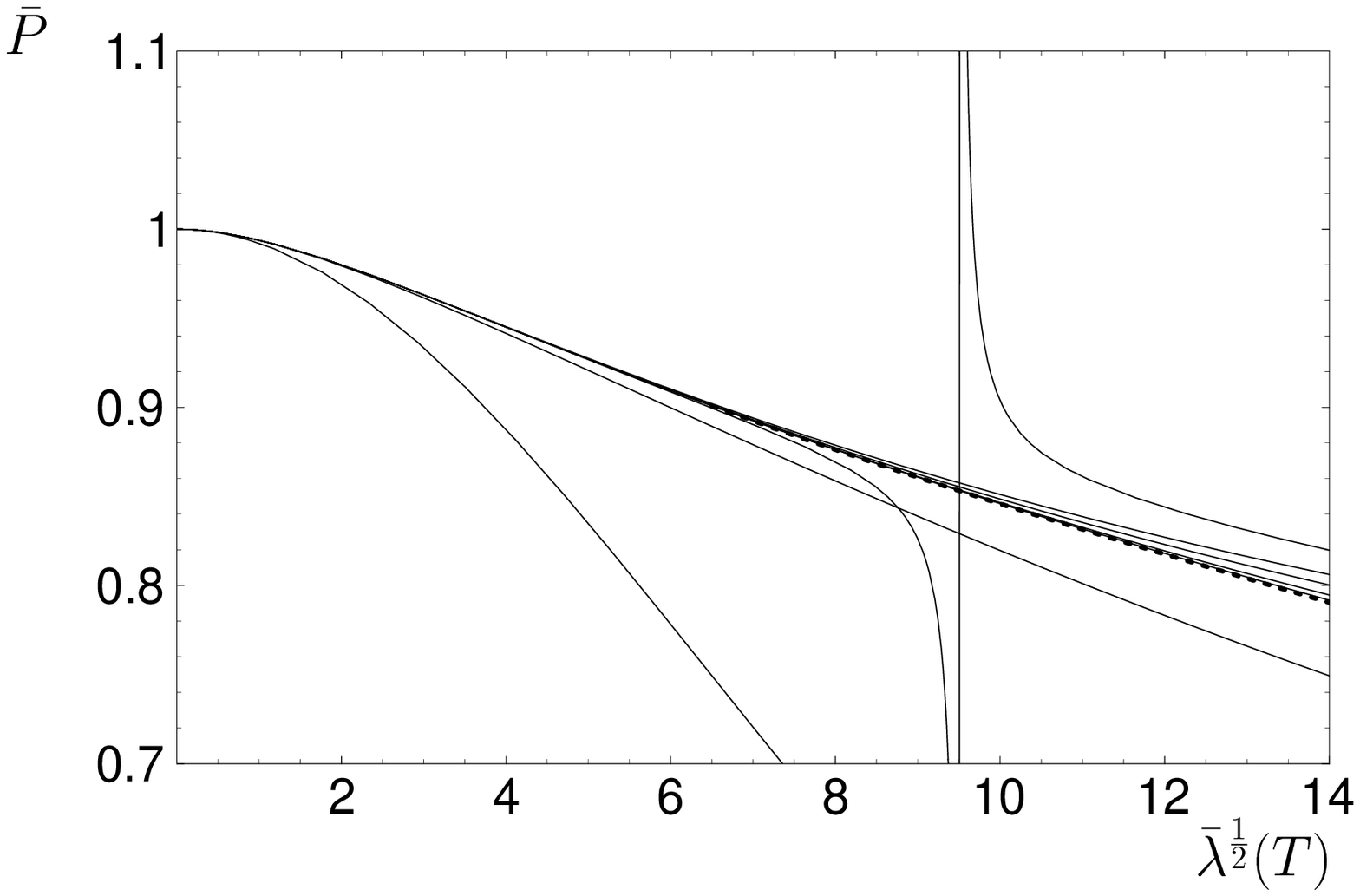}}
}
\bigskip
\moveright1cm\vbox{\hsize=14cm
Figure 8: As in figure 6b, but with the perturbative power series replaced
by Pad\'e approximants $[0,2],\;[1,2],\;[2,2],\;[2,3],\ldots,[4,4]$. 
They quickly
converge to the exact result with the exception of $[3,3]$, which has
a pole beyond which this approximant seems off by a constant.
}
\endinsert

\topinsert
\centerline{{\epsfxsize=95truemm
\epsfbox{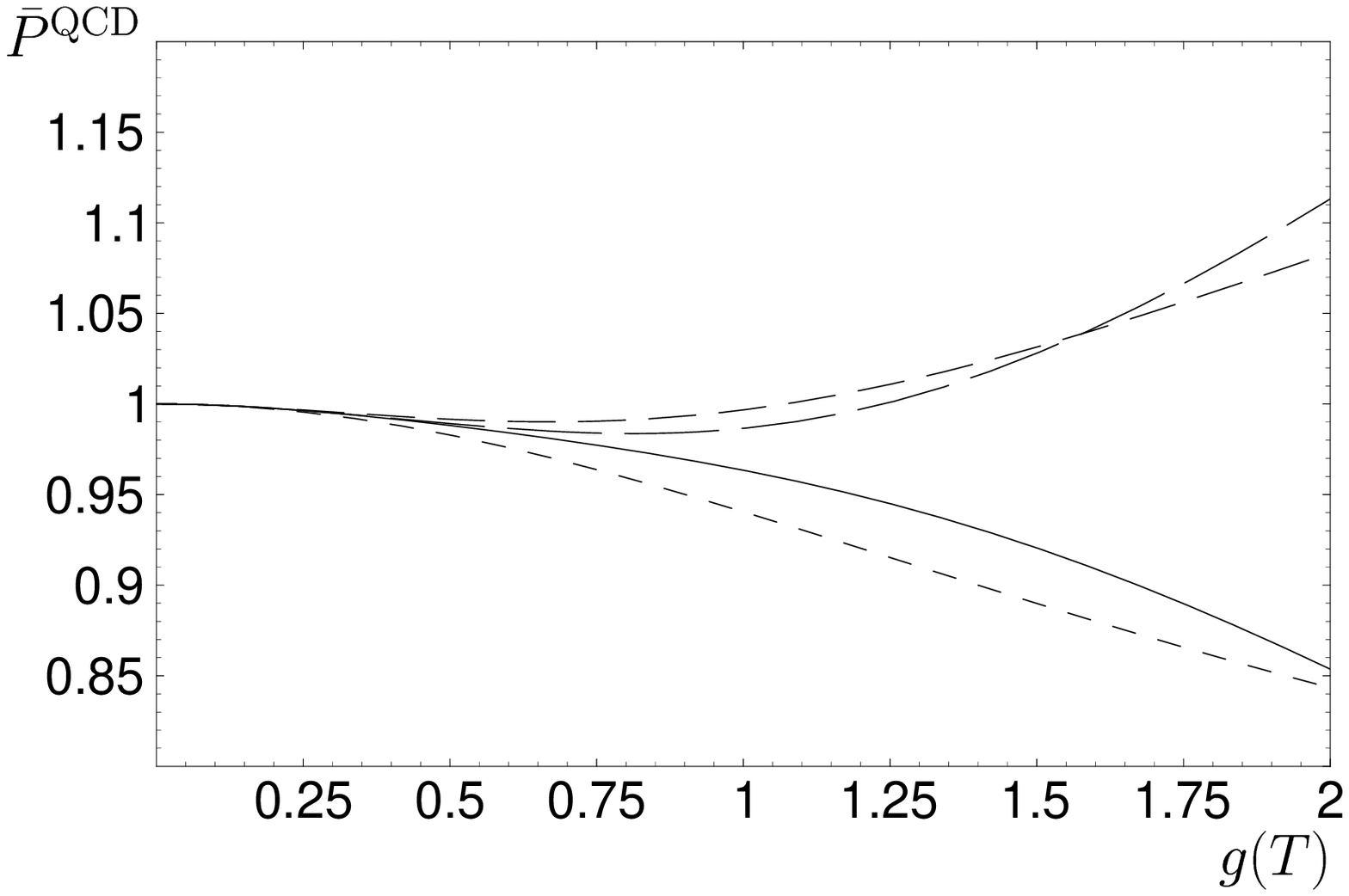}}}
\centerline{(a)}
\medskip
\centerline{\epsfxsize=95truemm
\epsfbox{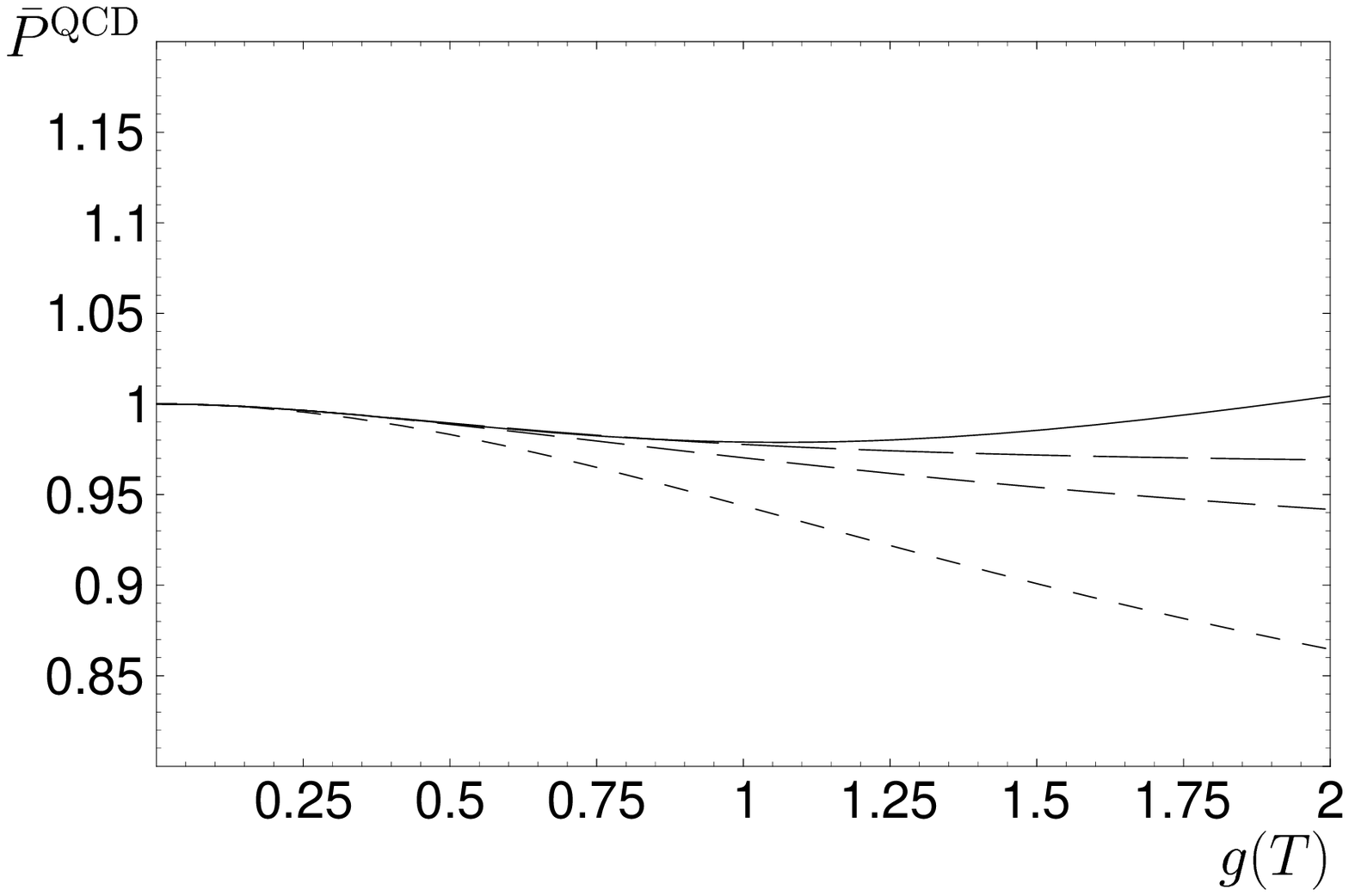}}
\centerline{(b)}
\medskip
\moveright1cm\vbox{\hsize=14cm
Figure 9: a) The perturbative results for the pressure of QCD($N_f=3$)
up to order $g^5$ (full line). Short, medium, and long dashes give the results
up to $g^2$, $g^3$, and $g^4$, respectively. b) The corresponding
Pad\'e approximants $[0,2]$, $[1,2]$, $[2,2]$, and $[3,2]$.}
\endinsert

\topinsert
\centerline {
{\epsfxsize=95truemm
\epsfbox{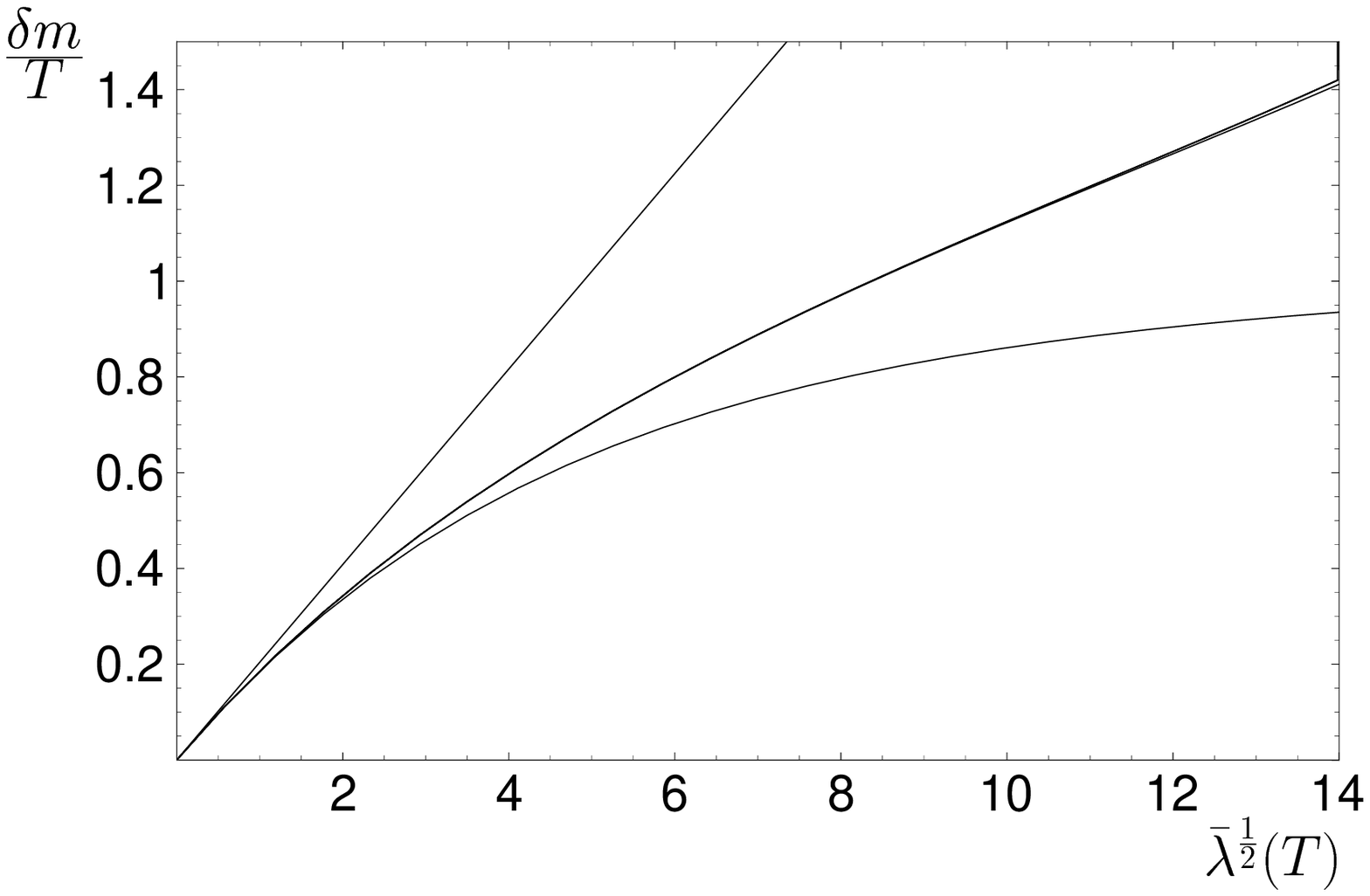}}
}
\centerline{(a)}
\medskip
\centerline{\epsfxsize=95truemm
\epsfbox{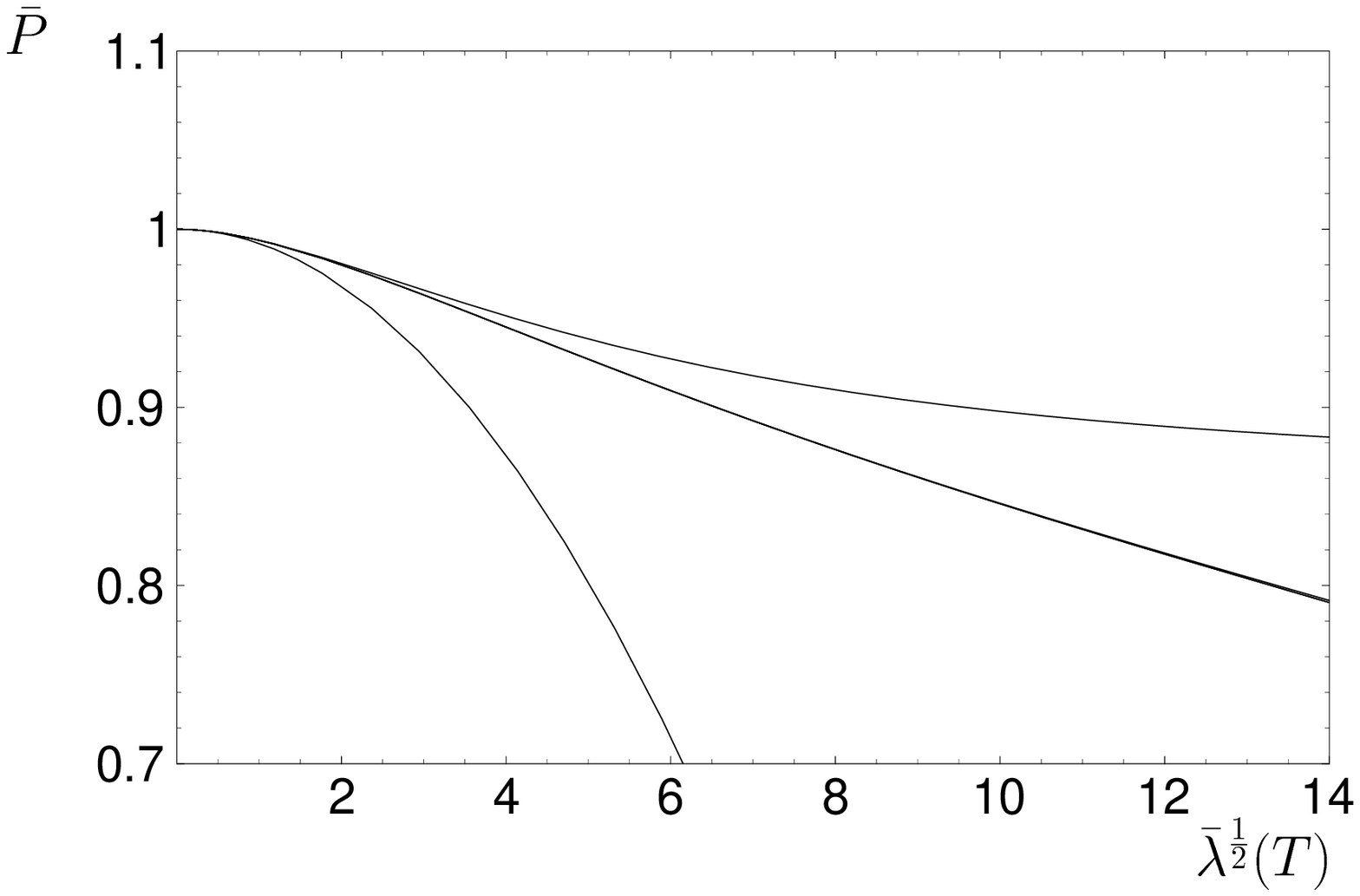}}
\centerline{(b)}
\medskip
\moveright1cm\vbox{\hsize=14cm
Figure 10: a) Solutions of the thermal mass equation when the latter
is truncated at leading order in $\delta m/T$ (top curve) and 
next-to-leading order (bottom curve). 
All higher approximations are virtually
indistinguishable from the exact solution,
and they collectively form the middle line, which broadens
only at the highest values of $\bar\la$.
b) Analogously for the pressure.
}
\endinsert

A rather simple reorganization of perturbation theory is brought about
by replacing truncated power series expansions by perturbatively
equivalent Pad\'e approximants\defref{\Pade}{G A Baker, 
{\it Essentials of Pad\'e Approximants}, Academic Press (1975)},
which amounts to replacing
$$
F_n=c_0+c_1 g^1+\ldots c_n g^n \to 
F_{[p,n-p]}=
{c_0 + a_1 g^1 + \ldots + a_p g^p \over
1+ a_{p+1} g^1 + \ldots + a_n g^{n-p}}+O(g^{n+1})
\eqno(5.9)$$
In reference \defref{\KP}{B Kastening, hep-ph/9708219},
this possibility has been studied in the context of thermal perturbation
theory by testing for an unphysical dependence on the
renormalization scale. Our results allow us to
investigate the quality of Pad\'e approximants by comparing directly with
truncated power series and the exact result.

In figure 8 we have replaced the various $n$th-order results that
gave rise to the curves displayed in figure 6b by $[n/2,n/2]$ Pad\'e
approximants up to $n=8$
(for odd $n$ we rounded off the second entry at the expense
of the first) and using $\bar\mu=T$. 
It turns out that we find a spectacular improvement
of convergence up to really high values of the coupling which
in our theory is bounded by the requirement to be sufficiently below
the critical value, $\bar\la^{1/2}(T)\ll18$. The lower 
line in figure 8 is the Pad\'e approximant $[0,2]$ ($[1,1]$ does not exist), 
which is only marginally better than its perturbative analogue $[2,0]$,
but already $[1,2]$ (given by the next line upwards) is a
very good approximation.
The higher approximants approach the exact result from above and are
extremely accurate with the exception of $[3,3]$ which
has a pole at $\bar\la^{1/2}\approx9.5$. 
For $\bar\la^{1/2}$ sufficiently smaller than that, this approximant
is in fact quite good, 
but after the pole has been encountered it seems to be
off by a constant\footnote{*}{In reference \ref{\KP}
poles in Pad\'e approximants were simply subtracted. Our observations
imply that this works nicely only when the coupling is such that one is below
the point where a pole arises.} and its quality is inferior
to most of the lower-order approximants. 

Curiously enough, with the choice $\bar\mu=2 \pi T$
which has led to more rapid convergence in the case of truncated power series,
the Pad\'e approximants are not as good as in figure 8, although they
are still a great improvement for $\bar\la^{1/2}<8$. 
What happens is that all of the
approximants run into poles in the range of $\bar\lambda$ considered.
The same holds true for larger $\bar\mu$, whereas for $\bar\mu \ll T$
the quality of the Pad\'e approximations decreases, too, but not as
rapidly.

All in all it appears that Pad\'e approximants can give vast
improvements of a truncated power series expansion unless the rational
functions used as approximants develop poles. While there is no
real theoretical explanation for the superiority of
Pad\'e approximants, it seems that their main advantage is that
(in the absence of poles)
the latter do not blow up at larger coupling
as quickly as the corresponding truncated
power series inevitably do. Since the exact result behaves rather
unspectacularly, the odds seem to be in favour of Pad\'e approximants.

Unfortunately, in QCD Pad\'e approximants turn out to lead to less
impressive improvements\ref{\KP}. In figure 9a the perturbative result
for QCD with 3 flavours is given, which shows that (resummed) perturbation
theory is useful only up to $g(T)\approx 1/2$. But a real quark-gluon plasma
as one hopes to produce in heavy-ion collisions has rather $g(T)\approx2$,
where the perturbative results are completely inconclusive.
The corresponding Pad\'e approximants are rendered in figure 9b. They
seem to give some improvement of convergence, extending the allowed
range of coupling to perhaps $g(T)\approx1$, but this appears to break down
before reaching $g(T)=2$.

So ultimately one would have to find a different expansion scheme
that does not involve truncated series in the coupling if one wants to cover
more-strongly-coupled theories. 
Recently an interesting attempt towards an alternative perturbative scheme has
been made in the example of a scalar theory
in reference \defref\KPP{F Karsch, A Patk\'os and P Petreczky, 
Phys Lett B401 (1997) 69} 
using the numerical solution of an approximation to the gap equation in a loop
expansion of the pressure,
which however required an ad hoc treatment
of uncancelled ultraviolet divergences. 

In this context we observe that while
the perturbative results are satisfactory only
for small coupling, the high-temperature series (5.4) and (5.6) 
have excellent convergence properties, with
convergence radius $\d m/T = 2\pi$.
In our model,
the exact result for the thermal mass remains sufficiently
smaller than $2\pi T$ even for extremely large coupling $\bar\lambda$. 
Only when we re-expanded
in $\bar\lambda$ did 
the convergence become bad. 
In order to highlight this,
figure 10a shows the solutions to the thermal mass 
equation when it is truncated
at the leading  or next-to-leading term  in the expansion in
$\d m/T$, rather than being expanded in $\bar\lambda$. 
Including the next power of 
$\d m^2/T^2$, together with the logarithm, turns out to produce a result
which is virtually indistinguishable from the exact one
for the whole range of
$\bar\lambda$ bounded by
the requirement that the tachyon mass remains larger than anything else.
The same holds true for the pressure (figure 10b), with even smaller
deviations from the exact result.

In the particularly interesting case of 
gauge theories it is crucial to have
a consistent expansion scheme with a well-defined expansion parameter
in order to be able to retain gauge
fixing independence. If in QCD it would be possible to reorganize
perturbation theory as a series in $\delta m/T$ rather than $g$,
figure~10 suggests that this should lead to dramatic improvements.

{\sl~
\bigskip
This research is supported in part by the EU Programme ``Human Capital
and Mobility", Network ``Physics at High Energy Colliders'', contract
CHRX-CT93-0357 (DG 12 COMA), by the Austrian 
``Fonds zur F\"orderung der wissenschaftlichen
Forschung (FWF)'', project no. P10063-PHY, by the Jubil\"aumsfonds der
\"Osterreichischen Nationalbank, project no. 5986, by
the Leverhulme Trust and by PPARC.
We are much indebted to Andras Patkos for pointing out a mistake
in an earlier version of this paper.}

\vfill\eject

\medskip\immediate\closeout\rfile\writestoppt
\baselineskip=14pt{{\bf References}}\bigskip{\frenchspacing%
\parindent=20pt\escapechar=` \input refs.tmp\bigskip}\nonfrenchspacing
\bye